\documentclass[10pt,preprint]{aastex}

\newcommand\TA{\tablenotemark{a}}
\newcommand\TB{\tablenotemark{b}}
\newcommand\TC{\tablenotemark{c}}

\newcommand\mcnd{\multicolumn{2}{c}{\nodata}}

\shorttitle{30~Doradus Spectrophotometry}
\shortauthors{Peimbert}

\begin{document}

\title{The Chemical Composition of the 30~Doradus Nebula Derived
from VLT Echelle Spectrophotometry\footnotemark{}}

\author {Antonio Peimbert}
\affil {Instituto de Astronom\'\i a, UNAM,
Apdo. Postal 70-264, M\'exico 04510 D.F., M\'exico} 
\email{antonio@astroscu.unam.mx}

\begin{abstract}
Echelle spectrophotometry of the 30~Doradus nebula in the LMC is presented. The
data consists of VLT UVES observations in the 3100 to 10350 \AA\ range. The
intensities of 366 emission lines have been measured, including 269 identified
permitted lines of H$^{0}$, He$^{0}$, C$^{0}$, C$^{+}$, N$^{+}$, N$^{++}$,
O$^{0}$, O$^{+}$, Ne$^{0}$, Ne$^{+}$, S$^{+}$, S$^{++}$, Si$^{0}$, Si$^{+}$,
Si$^{++}$, Ar$^{+}$, and Mg$^{+}$; many of them are produced by recombination
only while others mainly by fluorescence.  Electron temperatures and densities
have been determined using different line intensity ratios.  The He$^+$,
C$^{++}$, O$^+$, and O$^{++}$ ionic abundances have been derived from
recombination lines, these abundances are almost independent of the temperature
structure of the nebula. Alternatively abundances from collisionally excited
lines have been derived for a large number of ions of different elements, these
abundances depend strongly on the temperature structure. Accurate $t^2$ values
have been derived from the Balmer continuum, and by comparing the C$^{++}$,
O$^+$, and O$^{++}$ ionic abundances obtained from collisionally excited and
recombination lines. The chemical composition of 30~Doradus is compared with
those of Galactic and extragalactic \ion{H}{2} regions. The values of $\Delta
Y$/$\Delta O$, $\Delta Y$/$\Delta Z$, and $Y_p$ are also discussed.
\end {abstract}


\section{Introduction}

\footnotetext{Based on observations collected at the European Southern
Observatory, Chile, proposal number ESO 68.C-0149(A).}

The determination of the chemical composition of \ion{H}{2} regions has been
paramount for the study of the chemical evolution of galaxies and for the
determination of the primordial helium abundance, $Y_p$. In recent times the
determination of atomic data of higher accuracy and the detection of fainter
emission lines with the use of echelle spectrophotometry have permitted to
derive more accurate physical conditions for Galactic \ion{H}{2} regions
\citep[e.g.,][] {est98,est99a,est99b}. For these reasons it was decided to
carry out echelle spectrophotometry of 30~Doradus.

Due to its proximity, its high angular dimensions, and its high surface
brightness 30~Doradus, NGC~2070, is the most spectacular extragalactic
\ion{H}{2} region and thus it has been the subject of many spectrophotometric
studies
\citep*[e.g.,][]{pei74,all74,duf75,pag78,boe80,duf82,sha83,mat85,ros87,ver02,tsa02}.

The main aim of this paper is to make a new determination of the chemical
abundances of 30~Doradus including the following improvements over previous
determinations: the consideration of the temperature structure that affects the
helium and heavy elements abundance determinations, the derivation of the O and
C abundances from recombination line intensities of very high accuracy, the
consideration of the collisional excitation of the triplet \ion{He}{1} lines
from the $2^3$S level by determining the electron density from many line
intensity ratios, and the study the $2^3$S level optical depth effects on the
intensity of the triplet lines by observing a large number of singlet and
triplet lines of \ion{He}{1}.

In sections 2 and 3 the observations and the reduction procedure are
described. In section 4 temperatures and densities are derived from eight and
six different methods respectively; also in this section, four independent
values of the mean square temperature fluctuation, $t^2$, are determined by
combining the electron temperatures.  In section 5 ionic abundances are
determined based on recombination lines that are almost independent of the
temperature structure, and ionic abundances based on ratios of collisionally
excited lines to recombination lines that do depend on the temperature
structure of the nebula.  In section 6 the total abundances are determined and
compared with those of NGC~346 (the most luminous \ion{H}{2} region in the
SMC), the Orion nebula, M17, and the Sun.  In section 7 $\Delta Y/\Delta O$ and
$\Delta Y/\Delta Z$ are determined; these ratios are important restrictions for
the study of the chemical evolution of galaxies and for the determination of
the primordial helium abundance, $Y_p$.  Also in section 7 $Y_p$ is determined
based on the abundances of 30 Dor and a value of $\Delta Y/\Delta O$ derived
from the observations of other objects and from chemical evolution models of
irregular galaxies.

\section{Observations}

The observations were obtained with the Ultraviolet Visual Echelle
Spectrograph, UVES \citep{dod00}, at the VLT Kueyen Telescope in Chile. We
observed simultaneously with the red and blue arms in two settings, covering
the region from 3100 \AA\ to 10360 \AA ~(see Table~\ref{tobs}). The wavelength
regions 5783--5830 \AA\AA\ and 8540--8650 \AA\AA\ were not observed due to the
separation between the two CCDs used in the red arm. There were also two small
gaps that were not observed, 10084--10088 \AA\AA\ and 10252--10259 \AA\AA,
because the two redmost orders did not fit within the CCD.  In addition to the
long exposure spectra we took 60 second exposures for the four observed
wavelength ranges to check for possible saturation effects.

The slit was oriented east-west and the atmospheric dispersion corrector (ADC)
was used to keep the same observed region within the slit regardless of the air
mass value.  The slit width was set to 3.0" and the slit length was set to 10"
in the blue arm and to 12" in the red arm; the slit width was chosen to
maximize the S/N ratio of the emission lines and to maintain the required
resolution to separate most of the weak lines needed for this project. The FWHM
resolution for the 30~Dor lines at a given wavelength is given by $\Delta
\lambda \sim \lambda / 8800$. The reductions were made for an area of 3"
$\times$ 10" located 64" N and 60" E of HD 38268, these coordinates are similar
to those of 30 Dor II, position observed by \citet{pei74}, but the area
observed by us is considerably smaller. The slit was placed in a region of high
emission measure that does not include luminous stars.

The spectra were reduced using the IRAF\footnotemark{} echelle reduction
package, following the standard procedure of bias subtraction, aperture
extraction, flatfielding, wavelength calibration and flux calibration. For flux
calibration the standard star EG 247 was observed.

\footnotetext{IRAF is distributed by NOAO, which is operated by AURA,
under cooperative agreement with NSF.}

\section{Line Intensities and Reddening Correction}

Line intensities were measured integrating all the flux in the line between two
given limits and over a local continuum estimated by eye. In the few cases of
line-blending, the line flux of each individual line was derived from a
multiple Voigt profile fit procedure. All these measurements were carried out
with the {\tt splot} task of the IRAF package.

An initial reddening coefficient, $C$(H$\beta$), was determined by fitting the
observed $I$(H$\beta$)/$I$(H Balmer lines) ratios (with the exception of Balmer
$\alpha$) to the theoretical one computed by \citet{sto95} for $T_e$ = 9,000 K
and $N_e$ = 300 cm$^{-3}$, (see below) and assuming the extinction law of
\citet{sea79}. With this initial fit the the observed $I$(H$\alpha$)/
$I$(H$\beta$) and the $I$(Paschen lines)/ $I$(H$\beta$) ratios become larger
than predicted.  A similar situation prevails in the Orion Nebula
\citep[e.g.,][]{cos70,car88,gre94}.

A simultaneous fit to all the H lines was obtained by adopting the following
reddening law: it was assumed that the dust distribution is well represented by
Seaton's reddening law, but that part of the dust is in front of the nebula and
part is well mixed with the gas. A good fit for all the Paschen and Balmer
lines is obtained by adopting a $C$(H$\beta$) in front of the nebula of 0.31
dex and a $C$(H$\beta$) from the front to the back of the \ion{H}{2} region of
1.70 dex.  The effective $C$(H$\beta$) amounts to 0.92 dex and the effective
$R$ value to 4.2 (where $R = A_V/E(B-V)$ ), while Seaton's law provides an $R$
value of 3.2 for material in front of the source.

There are other assumptions that can produce the same or a similar extinction
law. For example if all the extinction occurs in front of the nebula but is not
uniform across the observing slit it is also possible to obtain effective $R$
values higher than 3.2 without changing the grain properties. Also it is
possible to have all the extinction between us and the \ion{H}{2} region caused
by grains with a different size distribution to that implied by Seaton's law.
Consequently the adopted extinction law is empirical and it is not the purpose
of this paper to estimate the relative proportion of the different causes that
produce an $R$ value higher than 3.2.

Since we avoided the brightest stars, the stellar continuum is very weak and
the equivalent width of H$\beta$ in emission amounts to 734 \AA, consequently
the correction due to underlying absorption of the H and He lines is negligible
and was not taken into account.

Table~\ref{tlines} presents the emission line intensities of 30 Dor.  The first
and second columns include the observed wavelength in the framework of the
solar standard of rest, $\lambda$, and the adopted laboratory wavelength,
$\lambda_0$. The third and fourth columns include the ion and the multiplet
number, or the Balmer, B, or Paschen, P, transitions for each line.  The fifth
and sixth columns include the observed flux relative to H$\beta$, $F(\lambda$),
and the flux corrected for reddening relative to to H$\beta$, $I(\lambda$). The
seventh column includes the fractional error (1$\sigma$) in the line
intensities.

366 distinct emission lines were measured; of them 43 are unidentified, 31 are
blends, and 292 are single lines. By combining the single and blended lines 356
line identifications were obtained; of them 269 are permitted, 85 forbidden and
2 semiforbidden.

Two of the lines presented in Table~\ref{tlines} correspond to a nebular
component in emission that originates in our Galaxy and that is not physically
associated with 30 Dor. This component has an emission measure in H$\alpha$
four orders of magnitude smaller than that of 30 Dor in the observed line
of sight. The H$\beta$ line of this component was also detected. Both Balmer
lines are blueshifted relative to 30 Dor by 252 km sec$^{-1}$. This nebular
Galactic component is not present in other wavelengths and it does not affect
the line intensities presented in Table~\ref{tlines}, with the possible
exception of 4098.24 \ion{O}{2}, that might be blended with the Galactic
H$\delta$ line.

\section{Physical Conditions}
\subsection{Temperatures and Densities}

The temperatures and densities presented in Table~\ref{ttemden} were derived
from the line intensities presented in Table~\ref{tlines}. Most of the
determinations were carried out based on the IRAF subroutines. The contribution
to the intensities of the $\lambda\lambda$ 7319, 7320, 7331, and 7332
[\ion{O}{2}] lines due to recombination was taken into account based on the
following equation:
\begin{equation}
I_R(7319+7320+7331+7332)/I({\rm H\beta})
= 9.36(T/10^4)^{0.44} \times {\rm{O}}^{++}/{\rm{H}}^+,
\end{equation}
\citep[see][]{liu00}. The [\ion{Cl}{3}] temperature was obtained from the
$I$(8500)/$I$(5518 + 5538) ratio and the computations by \citet{kee00}. The
Balmer continuum temperature was determined from the following equation:
\begin{equation}
T = 368 \times(1 + 0.259y^+ + 3.409y{^++})({\rm Bac/H11}) ~{\rm {K}},
\end{equation}
\citep[see][]{liu01} where Bac/H11 is in \AA$^{-1}$, $y^+$ = He$^+$/H$^+$, and
$y^{++}$ = He$^{++}$/H$^+$. Figure~\ref{fbalmer} presents the region near
the Balmer limit, where the Balmer continuum can be easily appreciated.

The [\ion{Fe}{3}] density was derived from the $I$(4986)/$I$(4658) ratio and
the computations by \citet{kee01}. In Table~\ref{tFepred} I compare the
observed $I(\lambda_{nm})$/$I$(4658) ratios for a set of [\ion{Fe}{3}] lines
with the predicted ratios by \citet{kee01} for $N_e$ = 316 cm$^{-3}$ and $T_e$
= 10,000 K.  I did not list in this table the [\ion{Fe}{3}] $\lambda\lambda$
4008, 4925 and 4931 lines because probably 4008 has a contribution from an
unidentified component, 4925 is blended with a permitted \ion{O}{2} line, and
4931 is blended with a forbidden [\ion{O}{3}] line. {From} Table~\ref{tFepred}
it can be seen that the observed and predicted line ratios are in excellent
agreement since the differences between them are of the order of their
estimated errors.

\subsection{Temperature variations}

To derive the ionic abundance ratios the average temperature, $T_0$, and the
mean square temperature fluctuation, $t^2$, were used. These quantities are
given by
\begin{equation}
T_0 (N_e, N_i) = \frac{\int T_e({\bf r}) N_e({\bf r}) N_i({\bf r}) dV}
{\int N_e({\bf r}) N_i({\bf r}) dV},
\end{equation}
and
\begin{equation}
t^2 = \frac{\int (T_e - T_0)^2 N_e N_i dV}{T_0^2 \int N_e N_i dV},
\end{equation}
respectively, where $N_e$ and $N_i$ are the electron and the ion densities of
the observed emission line and $V$ is the observed volume \citep{pei67}.

The \ion{H}{1} and \ion{He}{1} lines originate in the high and the low degree
of ionization zones and will be represented by $T_0$ and $t^2$.  I will divide
the other lines in two groups those that originate mainly in the high degree of
ionization zones: C$^{++}$, O$^{++}$, Ne$^{++}$, S$^{++}$, Cl$^{++}$,
Cl$^{+3}$, Ar$^{++}$, and Ar$^{+3}$, and those that originate mainly in the low
degree of ionization zones: N$^{+}$, O$^{+}$, S$^{+}$, Cl$^{+}$, and Fe$^{++}$.
These two groups will be represented by $T_{0,H}$, $T_{0,L}$, $t_H^2$, and
$t_L^2$.  I will also assume that $t_H^2 = t_L^2 \approx t^2$.

To determine $T_0$, $T_{0,H}$, $T_{0,L}$, and $t^2$ we need at least three
independent $T_e$ determinations that weigh differently the high and low
temperature zones and the fraction of the emissivity originating in the high
and low ionization zones, that for these observations corresponds to 85\% and
15\% respectively (see below).  For example, it is possible to combine the
temperature derived from the ratio of the [\ion{O}{3}] $\lambda\lambda$ 4363,
5007 lines, $T_{(4363/5007)}$, the temperature derived from the ratio of the
[\ion{N}{2}] $\lambda\lambda$ 5755, 6584 lines, $T_{(5755/6584)}$, and the
temperature derived from the ratio of the Balmer continuum to $I(H\beta)$,
$T_{({\rm Bac}/H\beta)}$, that are given by
\begin{equation}
T_{(4363/5007)} = T_{0,H} \left[ 1 + {\frac{1}{2}}\left({\frac{90800}{T_0}} - 3
\right) t_H^2\right],
\end{equation}
\begin{equation}
T_{(5755/6584)} = T_{0,L} \left[ 1 + {\frac{1}{2}}\left({\frac{69000}{T_0}} - 3
\right) t_L^2\right],
\end{equation}
and
\begin{equation}
T_{({\rm Bac}/{\rm H}\beta)} = T({\rm Bac}) = T_0 (1 - 1.70 t^2).
\end{equation}
 
Similar equations can be derived from the [\ion{O}{2}], [\ion{S}{2}],
[\ion{S}{3}], and [\ion{Ar}{3}] auroral to nebular line ratios. {From} the
abundances of O$^+$ and O$^{++}$ I estimate that 85\% of the H($\beta$)
emissivity comes from the regions of high degree of ionization and 15\% from
the regions of low degree of ionization. By combining the forbidden line
temperatures, $T$(FL) with the $T_{({\rm Bac}/H\beta)}$ temperature I obtain
the $t^2$ value presented in Table~\ref{tt2}.

Under the assumption that $T$ is constant along the line of sight it is
possible to derive the abundances of a given ion from a collisionally excited
line of an element $p$ times ionized or from a recombination line of the same
element $p-1$ times ionized.  In many objects the abundances derived from the
recombination lines are higher than those derived from the collisionally
excited lines, possibly indicating the presence of temperature variations along
the line of sight.

Taking into account that the abundances derived from the recombination lines
and the collisionally excited lines are the same it is possible to derive from
the ratio of a collisionally excited line to a recombination line a function of
$T_{0,H}$ and $t_H^2$, or $T_{0,L}$ and $t_L^2$. By combining this relation for
lines that originate mainly in regions of high degree of ionization with a
temperature determined from the ratio of two collisionally excited lines, like
$T_{(4363/5007)}$, it is also possible to derive $T_{0,H}$ and $t_H^2$.

Table~\ref{tt2} presents three $t^2$ values derived from the O$^+$(R/C),
O$^{++}$(R/C), and C$^{++}$(R/C) line intensity ratios.  The lines used were:
for O$^+$, the three lines of multiplet 1 of \ion{O}{1} (7773) together with
the [\ion{O}{2}] $\lambda 3727$ lines; for O$^{++}$, the eight lines of
multiplet 1 of \ion{O}{2} (4659) together with the [\ion{O}{3}] $\lambda 5007$
line; and, for C$^{++}$, the \ion{C}{2} line $\lambda$ 4267 of multiplet 6
together with the $\lambda\lambda$ 1906 and 1909 collisionally excited lines of
\ion{C}{3}. These relations were combined with the temperatures derived from
the ratios of forbidden lines for the low degree of ionization zones in the
first case, and with the temperatures derived from the ratios of forbidden
lines for the high degree of ionization zones in the second and third cases.

\section{Ionic Chemical Abundances}

\subsection{Helium ionic abundances}

To obtain He$^+$/H$^+$ values we need a set of effective recombination
coefficients for the He and H lines, the contribution due to collisional
excitation to the helium line intensities, and an estimate of the optical depth
effects for the helium lines. The recombination coefficients used were those by
\citet{sto95} for H, and those by \citet{smi96} and \citet{ben99} for He. The
collisional contribution was estimated from \citet{saw93} and
\citet{kin95}. The optical depth effects in the triplet lines were estimated
from the computations by \citet*{ben02}.

Table~\ref{thelium} presents the He$^+$/H$^+$ values obtained from the eleven
best observed helium lines for $t^2 = 0.033$. Considering the observational
errors, for no value of $\tau_{3889}$ it is possible to find agreement in the
He$^+$/H$^+$ values among the six triplet helium lines. The four more sensitive
lines to $\tau_{3889}$ are $\lambda\lambda$ 3188, 3889, 4713, and 7065, two of
them, $\lambda\lambda$ 4713 and 7065, increase in 
intensity with increasing $\tau_{3889}$ and
two of them, $\lambda\lambda$ 3188 and 3889, decrease.

Two acceptable solutions for $\tau_{3889}$ have been obtained. By excluding
$\lambda\lambda$ 3188 and 3889, $\tau_{3889}$ = 4.4 is found; alternatively, by
excluding $\lambda\lambda$ 4713 and 7065, $\tau_{3889}$ = 10.5 is found.  This
result indicates that the computations for spherical geometry by \citet{ben02}
do not apply to the observed region of 30 Dor. This region is part of a bright
shell probably longer along the line of sight than in the plane of the sky,
obviously a non spherical object.

A fraction of the $np - 2s$ photons is converted into $\lambda\lambda$ 4713,
7065, 4471, and 5876 photons; due to geometrical effects a smaller fraction
than that expected, under the sphericaly simetric case, is sent in our
direction. I consider that it is a good approximation to assume that the
increase in the $\lambda\lambda$ 4713, 7065, 4471, and 5876 line intensities
corresponds to that predicted by the same $\tau_{3889}$ value and that the
decrease in the $\lambda\lambda$ 3188 and 3889 lines corresponds to another
$\tau_{3889}$ value. Consequently by averaging the values of the 9 helium lines
for $\tau_{3889}$ = 4.4 (excluding $\lambda\lambda$ 3188 and 3889) a value of
He$^+$/H$^+$ = 0.08470 $\pm 0.00068$ is obtained.

The five singlet lines, which are not affected by the 
$\tau_{3889}$ effect, yield
He$^+$/H$^+$ = 0.08448 $\pm 0.00099$ in excellent agreement with the value
derived from the nine helium lines and the discussion presented above.

{From} the observed spectra it is found that the $I$(4686)/$I$(H$\beta$) value
is smaller than 3.5~$\times~10^{-5}$, which together with the recombination
coefficients by \citet{bro71} imply that $N$(He$^{++}$)/$N$(H$^+$) is smaller
than 2.9~$\times~10^{-6}$.

\subsection{C and O ionic abundances from recombination lines}

The C$^{++}$ abundance was derived from the $\lambda 4267$ \AA~line of
\ion{C}{2} and the effective recombination coefficients computed by
\citet{dav00} for Case A and $T$ = 10,000 K.

The O$^+$ abundance was derived from the $\lambda \lambda$ 7771.96, and 7775.40
lines of \ion{O}{1} and the effective recombination coefficient for the
multiplet computed by \citet{peq91}. The third line of the multiplet, $\lambda$
7774.18, was partially blended with a telluric line in emission, consequently
it was not possible to measure its intensity, it was assumed that the three
lines of the multiplet are in LS coupling and consequently that $I(7774.18) =
I(7771.96 + 7775.40)/2$.

The O$^{++}$ abundance was derived from the eight lines of multiplet 1 of
\ion{O}{2} (see Figure~\ref{foii}) together with the effective recombination
coefficient for the multiplet computed by \citet{sto94} under the assumption of
Case B for $T_e = 10,000$ K and $N_e$ = 300 cm$^{-3}$.  The result is almost
independent of the case assumed, the difference in the O$^{++}$/H$^+$ value
between Case A and Case B is smaller than 4\%. It was found that the \ion{O}{2}
lines of multiplet 1 are in Case B based on the observed intensities of
multiplets 19, 2, and 28 of \ion{O}{2} that are strongly case sensitive;
\citet{pei93y} also found that the \ion{O}{2} lines in the Orion nebula are in
Case B. The line intensity ratios within multiplet 1 do not follow the LS
coupling predictions. Figure~\ref{foii} provides an excelent visual reference
to estimate the quality of the data, it includes 2 lines four orders of
magnitude fainter than H$\beta$ and also shows that lines separated by 2\AA\
are completely resolved.

\subsection{Ionic abundances from collisionally excited lines}

With the exception of C$^{++}$/H$^+$ and Fe$^{++}$/H$^+$ all the other values
presented in Table~\ref{tcl} for $t^2 = 0.00$ were derived with the IRAF task
{\tt abund}, using only the low- and medium-ionization zones. The low and
medium ionization zones of IRAF correspond to the low and high ionization zones
of this paper.

The C$^{++}$/H$^+$ value for $t^2 = 0.00$ was derived from the collisionally
excited lines of \ion{C}{3} $\lambda \lambda$ 1906 and 1909 by \citet{duf82}
and \citet{gar95}.  I consider this procedure valid because the O degree of
ionization derived here is in excellent agreement with theirs (O$^{++}$/O
equal to 85\% and 83\% respectively).

The Fe$^{++}$/H$^+$ value for $t^2 = 0.00$ was derived from the atomic data by
\citet{nah96} and \citet{zha96}. I did not determine the Fe$^+$/H$^+$
abundances because the observed [\ion{Fe}{2}] lines are produced by collisions
and by non negligible radiative processes difficult to estimate \citep{rod99}.

To derive the abundances for $t_H^2 = t_L^2 \approx t^2 = 0.033$ I used the
abundances for $t^2 = 0.00$ and the formulation for $t^2 > 0.00$ presented by
\citet{pei69}. To derive abundances for other $t^2$ values it is possible to
interpolate or to extrapolate the values presented in Table~\ref{tcl}.

\section{Total Abundances}

Table~\ref{ttotab} presents the total abundances of 30~Doradus for $t_L^2 =
t_H^2 \approx t^2 = 0.033\approx t^2$.  To derive the total gaseous abundances
the set of equations presented below was used, where the ionization correction
factors, $ICF$'s, correct for the unseen ionization stages.

The total He/H value is given by:
\begin{eqnarray}
\nonumber
\frac{N ({\rm He})}{N ({\rm H})} & = & \frac {N({\rm He}^0) + N({\rm He}^+) + 
N({\rm He}^{++})}{N({\rm H}^0) + N({\rm H}^+)},\\
& = & ICF({\rm He}) \frac {N({\rm He}^+) + N({\rm He}^{++})}
{N({\rm H}^+)}.                 
\end{eqnarray}
The He$^{++}$/H$^+$ ratio is completely negligible (see previous section).  

In objects of low degree of ionization the presence of neutral helium inside
the \ion{H}{2} region is important and $ICF$(He) becomes larger than 1. To
study this problem \citet[see also \citealt{pag92}]{vil88} defined a radiation
softness parameter given by
\begin{equation}
\zeta = \frac {N({\rm O}^+)N({\rm S}^{++})}{N({\rm S}^+)N({\rm O}^{++})};
\end{equation}                     
for large values of $\zeta$ the amount of neutral helium is significant, while
for low values of $\zeta$ it is negligible, where the critical value is around
8. {From} the previous equation and the values in Table~\ref{tcl} it is found
that $\zeta$ = 2.82 which indicates that the amount of He$^0$ inside the H$^+$
region is negligible.

On the other hand, for ionization bounded objects of very high degree of
ionization the amount of H$^0$ inside the He$^+$ Str\"omgren sphere becomes
significant and the $ICF$(He) can become smaller than 1. This possibility was
firstly mentioned by \citet{shi74} and studied extensively by \citet{arm99},
\citet{vie00}, \citet*{bal00} and \citet{sau02}.

According to \citet{bal00} for [\ion{O}{3}]$\lambda 5007$/[\ion{O}{1}]$\lambda
6300 \geq 300$, the $ICF$(He) becomes very close to unity; from
Table~\ref{tlines} it is found that $I(5007)/I(6300) = 548$. Consequently I
conclude that the amount of H$^0$ inside the observed \ion{H}{2} region is
negligible and in what follows I will adopt an $ICF$(He) = 1.000.

The gaseous abundances for O, N, and Ne were obtained from the following
equations \citep{pei69}
\begin{equation}
 \frac{N(\rm O)}{N(\rm H)} =
             \frac{N({\rm O^+})+N(\rm O^{++})}{N(\rm H^+)},
\end{equation}
\begin{equation}
 \frac{N(\rm N)}{N(\rm H)} =
             \left( \frac{N({\rm O^+})+N(\rm O^{++})}{N(\rm O^+)} \right)
             \frac{N(\rm N^+)}{N(\rm H^+)},
\end{equation}
and
\begin{equation}
\frac{N(\rm Ne)}{N(\rm H)} =
             \left( \frac{N({\rm O^+})+N(\rm O^{++})}{N(\rm O^{++})} \right)
             \frac{N(\rm Ne^{++})}{N(\rm H^+)}.
\end{equation}

To obtain the total O/H gaseous abundance the O$^+$/H$^+$ and O$^{++}$/H$^+$
values presented in Tables~\ref{trl} and~\ref{tcl} were weighed according to
their observational errors.  To obtain the total O abundances a correction of
0.08~dex was adopted to take into account the fraction of O tied up in dust
grains, this fraction was estimated from the Mg/O, Si/O, and Fe/O values
derived for the Orion nebula \citep{est98}.
 
To obtain the C gaseous abundance the following equation was adopted
\begin{equation}
\label{ecarbon}
\frac{N(\rm C)}{N(\rm H)} = ICF({\rm C}) \frac{N({\rm C^{++}})}{N(\rm H^+)},
\end{equation}
where the C$^{++}$/H$^+$ gaseous abundance was obtained by weighing the
C$^{++}$/H$^+$ values presented in Tables~\ref{trl} and~\ref{tcl} according to
their observational errors, the $ICF$(C) value was obtained from
\citet{gar95}. Following \citet{est98} 0.10 dex were added to the total C/H
gaseous value to take into account the fraction of C atoms embedded in dust
grains.

The gaseous abundances of S, Cl, Ar and Fe were obtained from the following
equations:
\begin{equation}
\label{esulphur}
\frac{N(\rm S)}{N(\rm H)} = ICF({\rm S}) \frac{N({\rm S^+}) + N(\rm
S^{++})}{N(\rm H^+)},
\end{equation}
\begin{equation}
\frac{N(\rm Cl)}{N(\rm H)} = \frac{N({\rm Cl^+}) + N({\rm Cl^{++}}) + N(\rm Cl^{3+})}{N(\rm H^+)},
\end{equation}
\begin{equation}
\frac{N(\rm Ar)}{N(\rm H)} = ICF({\rm Ar}) \frac{N({\rm Ar^{++}}) + N(\rm
Ar^{3+})}{N(\rm H^+)},
\end{equation}
and
\begin{equation}
\label{eiron}
\frac{N(\rm Fe)}{N(\rm H)} = ICF({\rm Fe}) \frac{N({\rm Fe^{++}})}{N(\rm H^+)}.
\end{equation}
The $ICF$(S) value was estimated from the models by \citet{gar89}. The
$ICF$(Fe) value was estimated from the models of NGC~346 by \citet*{rel02} and
amounts to $8~\pm 2$.  The $ICF$(Ar) includes the Ar$^+$/H$^+$ contribution and
according to \citet{liu00} can be approximated by
\begin{equation}
ICF({\rm Ar}) =\left( 1- \frac{N(\rm O^+)}{N(\rm O)}\right)^{-1}.
\end{equation}

\section{Discussion and Conclusions}

\subsection{ The \ion{H}{2} regions and the solar abundances}

Table~\ref{tta} presents the abundances of four very well observed \ion{H}{2}
regions: Orion and M17 in the Galaxy, NGC~346 in the SMC, and 30~Doradus in the
LMC. The Orion, M17, and 30~Doradus values include additions to the C and O
gaseous values of 0.1 dex and 0.08 dex respectively to take into account the
fraction of these atoms embedded in dust. For the same reason the NGC~346 C and
O values include additions of 0.05 dex and 0.04 dex respectively
\citep{rel02}. The \ion{H}{2} region abundances have been obtained adopting
values of $t^2$ larger than 0.00. Further arguments in favor of $t^2 > 0.00$
have been presented elsewhere \citep{pei02y,pei02a,pei02b}.

In addition Table~\ref{tta} presents also the solar photospheric values for C,
N, O, Ne, and Ar, and the solar abundances derived from meteoritic data for S,
Cl, and Fe. For the solar initial helium abundance the $Y_0$ by \citet{chr98}
was adopted, and not the photospheric one because, apparently, it has been
affected by settling.

The \ion{H}{2} region values for Ne/O, S/O, and Ar/O are in excellent agreement
with the solar values which implies that the production of these elements is
primary and due to massive stars, and that the assumptions involved in the two
types of abundance determinations are sound.

The \ion{H}{2} regions S/O and Cl/O abundances are in better agreement with the
solar meteoritic abundances than with the photospheric ones, the solar
photospheric values are S/O = -1.38 $\pm 0.11$ dex, and Cl/H = -3.21 $ \pm 0.3$
dex.

The 30~Doradus C/O value is intermediate between that of NGC~346 and those of
Orion, M17 and the Sun. The differences are significant and imply that even if
C is of primary origin part of it is due to intermediate mass stars and 
part is due to
massive stars, and that the C yield increases with the O/H ratio
\citep{gar95,gar99,car02}.

The 30~Doradus N/O value is intermediate between that of NGC~346 and those of
Orion, M17 and the Sun. The differences are significant and imply that part of
the N is of primary origin and part of secondary origin \citep[see][and
references therein]{hen00}.

The accuracies of the He/H abundances of 30~Doradus, NGC~346, and M17 are
higher than that of the Orion nebula because the first three objects have
$ICFs$(He) = 1.00, while the $ICF$(He) for the Orion nebula is larger than 1.00
and it is not well determined. Similarly the values for 30~Doradus, NGC~346,
and M17 are more accurate than the solar one because they are based on direct
determinations, while the solar value is obtained from models that depend on
the helium abundance in a more complex way.

\subsection{The $\Delta Y$/$\Delta O$ and $\Delta Y$/$\Delta Z$ ratios 
and chemical evolution}

The determination of the $\Delta Y$/$\Delta O$ and $\Delta Y$/$\Delta Z$ ratios
is crucial for the determination of $Y_p$, and
for constraining the models of galactic chemical evolution. The abundance
determinations of 30~Doradus are based on emission line intensities of high
quality, take into account the temperature structure of the nebula, and include
a very accurate He/H value because the degree of ionization of 30~Doradus is
relatively high implying an $ICF$(He) very close to unity.

To determine the hydrogen, helium, heavy elements, and oxygen abundance by
mass, $X$, $Y$, $Z$, and $O$, presented in Table~\ref{tta} I proceeded as
follows: for the Sun I adopted the initial helium abundance by mass ($Y_0$) of
\citet{chr98}, the $Z/X$ value derived from the C/H, N/H, O/H, Ne/H, and Ar/H
values presented in Table~\ref{tta}, and $0.56 \times O/X$ for the ratio of the
rest of the heavy elements to hydrogen, value obtained from the meteoritic
abundances by \citet{gre98}; for the \ion{H}{2} regions I adopted the He/H
value determined by the different observers, the C/H, N/H, O/H, Ne/H, and Ar/H
values presented in Table~\ref{tta}, and $0.56 \times O/X$ for the rest of the
heavy elements.

{From} Table~\ref{tdy/dz} it can be seen that the spread among the $\Delta
Y$/$\Delta O$ and $\Delta Y$/$\Delta Z$ values derived from the \ion{H}{2}
regions is smaller for the $t^2 > 0.00$ results than for the $t^2 = 0.00$
results, which is consistent with the idea that the $t^2 > 0.00$ values are
better. Moreover the theoretical computations for the chemical evolution of
irregular galaxies and for the solar vicinity predict values of $\Delta
Y$/$\Delta O$ in the 2.9 to 4.6 range with a representative value of $\Delta
Y$/$\Delta O$ = 3.5 $\pm 0.9$ \citep*[see][]{car95,car99,car00,chi97,pei00},
again in better agreement with the results from \ion{H}{2} regions under the
assumption that $t^2 > 0.00$.

\subsection{The primordial helium abundance}

To determine the $Y_p$ value from 30 Doradus it is necessary to estimate the
fraction of helium present in the interstellar medium produced by galactic
chemical evolution. For this purpose it was assumed that:
\begin{equation}
\label{DeltaO}
Y_p  =  Y({\rm 30~Dor}) - O({\rm 30~Dor}) \frac{\Delta Y}{\Delta O}.
\end{equation}
As in section 7.2 a $\Delta Y$/$\Delta O$ = 3.5 $\pm 0.9$ will be adopted,
which together with the $Y$(30~Dor) and $Z$(30~Dor) values of Table~\ref{tta}
yield $Y_p$ = 0.2345 $\pm$ 0.0047, where most of the error comes from the
adopted $\Delta Y$/$\Delta O$ value. This $Y_p$ value is in excellent agreement
with the value derived from NGC 346 \citep*{pei00}.  This agreement is due to
the similarity between the adopted $\Delta Y$/$\Delta O$ value and that derived
from 30 Doradus and NGC 346.
 
\acknowledgments I am grateful to Mar\'{\i}a Teresa Ruiz and Manuel Peimbert
for carrying out the observations presented in this paper and for fruitful
discussions, to C\'esar Esteban for his collaboration in the initial stages of
this project, and to Anabel Arrieta for her assistance in the reduction
process. It is also a pleasure to thank the Departamento de Astronom\'{\i}a de
la Universidad de Chile for its hospitality during a visit where part of this
work was made.

\clearpage

\begin{deluxetable}{l@{\hspace{48pt}}r@{--}lc} 
\tablecaption{Journal of Observations
\label{tobs}}
\tablewidth{0pt}
\tablehead{
\colhead{Date} & 
\multicolumn{2}{c}{$\lambda$ (\AA)} &
\colhead{Exp. Time (s)}}
\startdata
2002 March 12 & 3100 &3880  & 60, 3$\times$300 \\
2002 March 12 & 3730 &4990  & 60, 3$\times$900 \\
2002 March 12 & 4760 &6840  & 60, 3$\times$300 \\
2002 March 12 & 6600 &10360 & 60, 3$\times$900 \\
\enddata
\end{deluxetable} 

\clearpage

\begin{deluxetable}{r@{}lr@{}ll@{ }rr@{}lr@{}lr@{}l} 
\tablecaption{Line Intensities
\label{tlines}}
\tablewidth{0pt}
\tablehead{
\multicolumn{2}{c}{$\lambda$} &
\multicolumn{2}{c}{$\lambda_0$} &
\colhead{Ion} &
\colhead{Id.} &
\multicolumn{2}{c}{$F$} &
\multicolumn{2}{c}{$I$} &
\multicolumn{2}{c}{ err(\%)} }
\startdata
 3190&.198& 3187&.74& ~\ion{He}{1} &3        &   1&.450  &   2&.738  &  4&  \\
 3450&.269& 3447&.59& ~\ion{He}{1} &7        &   0&.1675 &   0&.2799 & 10&  \\
 3481&.996& 3478&.97& ~\ion{He}{1} &43       &   0&.0353 &   0&.0583 & 18&  \\
 3484&.850& 3481&.96& ~\ion{Ne}{2}?&6        &   0&.0515 &   0&.0848 & 15&  \\
 3490&.494& 3487&.73& ~\ion{He}{1} &42       &   0&.0997 &   0&.1638 & 10&  \\
 3501&.433& 3498&.66& ~\ion{He}{1} &40       &   0&.0601 &   0&.0984 & 15&  \\
 3515&.341& 3512&.52& ~\ion{He}{1} &38       &   0&.0487 &   0&.0791 & 15&  \\
 3533&.195& 3530&.50& ~\ion{He}{1} &36       &   0&.1431 &   0&.2307 & 10&  \\
 3557&.357& 3554&.42& ~\ion{He}{1} &34       &   0&.2049 &   0&.3268 &  7&  \\
 3590&.073& 3587&.28& ~\ion{He}{1} &31       &   0&.2161 &   0&.3403 &  7&  \\
 3593&.296& 3590&.45& ~\ion{Ne}{2}?&32       &   0&.0311 &   0&.0489 & 18&  \\
 3616&.427& 3613&.64& ~\ion{He}{1} &6        &   0&.2587 &   0&.4025 &  6&  \\
 3637&.054& 3634&.25& ~\ion{He}{1} &28       &   0&.2985 &   0&.4611 &  6&  \\
 3658&.755& 3656&.11& ~\ion{H}{1}  &H38      &   0&.0831 &   0&.1272 & 10&  \\
 3659&.391& 3656&.56& ~\ion{H}{1}  &H37      &   0&.0976 &   0&.1494 & 10&  \\
 3660&.073& 3657&.27& ~\ion{H}{1}  &H36      &   0&.0980 &   0&.1500 & 10&  \\
 3660&.731& 3657&.92& ~\ion{H}{1}  &H35      &   0&.1028 &   0&.1571 & 10&  \\
 3661&.461& 3658&.64& ~\ion{H}{1}  &H34      &   0&.1032 &   0&.1579 & 10&  \\
 3662&.211& 3659&.42& ~\ion{H}{1}  &H33      &   0&.1036 &   0&.1584 & 10&  \\
 3663&.099& 3660&.28& ~\ion{H}{1}  &H32      &   0&.1200 &   0&.1835 & 10&  \\
 3664&.047& 3661&.22& ~\ion{H}{1}  &H31      &   0&.1207 &   0&.1843 &  7&  \\
 3665&.074& 3662&.26& ~\ion{H}{1}  &H30      &   0&.1491 &   0&.2276 &  7&  \\
 3666&.199& 3663&.40& ~\ion{H}{1}  &H29      &   0&.1651 &   0&.2520 &  7&  \\
 3667&.483& 3664&.68& ~\ion{H}{1}  &H28      &   0&.1976 &   0&.3016 &  7&  \\
 3668&.948& 3666&.10& ~\ion{H}{1}  &H27      &   0&.2001 &   0&.3050 &  7&  \\
 3670&.534& 3667&.68& ~\ion{H}{1}  &H26      &   0&.2262 &   0&.3447 &  6&  \\
 3672&.312& 3669&.46& ~\ion{H}{1}  &H25      &   0&.2396 &   0&.3647 &  6&  \\
 3674&.295& 3671&.48& ~\ion{H}{1}  &H24      &   0&.2698 &   0&.4106 &  6&  \\
 3676&.585& 3673&.76& ~\ion{H}{1}  &H23      &   0&.2837 &   0&.4312 &  6&  \\
 3679&.201& 3676&.36& ~\ion{H}{1}  &H22      &   0&.3466 &   0&.5259 &  5&  \\
 3682&.188& 3679&.35& ~\ion{H}{1}  &H21      &   0&.3814 &   0&.5789 &  5&  \\
 3685&.655& 3682&.81& ~\ion{H}{1}  &H20      &   0&.4186 &   0&.6343 &  5&  \\
 3689&.688& 3686&.83& ~\ion{H}{1}  &H19      &   0&.5084 &   0&.7693 &  4&  \\
 3694&.408& 3691&.55& ~\ion{H}{1}  &H18      &   0&.5793 &   0&.8739 &  4&  \\
 3700&.000& 3697&.15& ~\ion{H}{1}  &H17      &   0&.7198 &   1&.084  &  4&  \\
 3706&.714& 3703&.85& ~\ion{H}{1}  &H16      &   0&.8483 &   1&.274  &  3&  \\
 3707&.874& 3705&.02& ~\ion{He}{1} &25       &   0&.4437 &   0&.6664 &  5&  \\
 3714&.838& 3711&.97& ~\ion{H}{1}  &H15      &   1&.033  &   1&.547  &  3&  \\
 3724&.702& 3721&.94& ~\ion{H}{1}  &H14      &   1&.857  &   2&.772  &  2&.5\\
 \mcnd    & 3721&.94& [\ion{S}{3}] &F2       &   \mcnd   &   \mcnd   & \mcnd\\
 3728&.942& 3726&.03& [\ion{O}{2}] &F1       &  40&.07   &  59&.74   &  1&.0\\
 3731&.698& 3728&.82& [\ion{O}{2}] &F1       &  42&.81   &  63&.73   &  1&.0\\
 3737&.257& 3734&.37& ~\ion{H}{1}  &H13      &   1&.446  &   2&.147  &  2&.5\\
 3753&.046& 3750&.15& ~\ion{H}{1}  &H12      &   1&.999  &   2&.954  &  2&.5\\
 3773&.547& 3770&.63& ~\ion{H}{1}  &H11      &   2&.611  &   3&.825  &  2&.0\\
 3792&.406& 3789&.43& ~\ion{C}{3}? &         &   0&.0430 &   0&.0626 & 12&  \\
 3800&.839& 3797&.8 & [\ion{S}{3}] &F2       &   3&.627  &   5&.264  &  2&.0\\
 \mcnd    & 3797&.92& ~\ion{H}{1}  &H10      &   \mcnd   &   \mcnd   & \mcnd\\
 3808&.856& 3805&.74& ~\ion{He}{1} &58       &   0&.0750 &   0&.1085 & 10&  \\
 3819&.474& 3816&.75& ~\ion{O}{3}? &         &   0&.0264 &   0&.0381 & 15&  \\
 3822&.593& 3819&.62& ~\ion{He}{1} &22       &   0&.8030 &   1&.157  &  3&  \\
 3836&.530& 3833&.57& ~\ion{He}{1} &62       &   0&.0239 &   0&.0342 & 15&  \\
 3838&.355& 3835&.39& ~\ion{H}{1}  &H9       &   5&.163  &   7&.394  &  1&.5\\
 3841&.055& 3838&.09& ~\ion{He}{1} &61       &   0&.0198 &   0&.0283 & 20&  \\
 3855&.562& 3852&.39& ~\ion{O}{2}  &177\TA   &   0&.0213 &   0&.0303 & 20&  \\
 3859&.033& 3856&.02& ~\ion{S}{3}  &1        &   0&.0602 &   0&.0856 & 10&  \\
 \mcnd    & 3856&.13& ~\ion{O}{2}  &12       &   \mcnd   &   \mcnd   & \mcnd\\
 3863&.030& 3860&.15& ~\ion{S}{2}? &41       &   0&.0127 &   0&.0181 & 25&  \\
 3865&.592& 3862&.60& ~\ion{S}{3}  &1        &   0&.0612 &   0&.0869 & 10&  \\
 3871&.759& 3868&.75& [\ion{Ne}{3}]&F1       &  23&.91   &  33&.84   &  1&.0\\
 3874&.825& 3871&.79& ~\ion{He}{1} &60       &   0&.0667 &   0&.0944 & 10&  \\
 3879&.016& 3875&.82& ~\ion{O}{2}  &13       &   0&.0132 &   0&.0187 & 20&  \\
 \mcnd    & 3876&.05& ~\ion{C}{2}  &33       &   \mcnd   &   \mcnd   & \mcnd\\
 3883&.183& 3880&.33& ~\ion{Ar}{2} &54       &   0&.0324 &   0&.0457 & 15&  \\
 3891&.947& 3888&.65& ~\ion{He}{1} &2        &  11&.14   &  15&.67   &  1&.2\\
 \mcnd    & 3889&.05& ~\ion{H}{1}  &H8       &   \mcnd   &   \mcnd   & \mcnd\\
 3904&.376& 3901&.35&              &         &   0&.0599 &   0&.0840 & 10&  \\
 3915&.934& 3912&.73& ~\ion{Mn}{1}?&         &   0&.0296 &   0&.0413 & 15&  \\
 3920&.198& 3917&.16&              &         &   0&.0113 &   0&.0158 & 25&  \\
 3922&.039& 3918&.98& ~\ion{C}{2}  &4        &   0&.0112 &   0&.0157 & 25&  \\
 3923&.706& 3920&.69& ~\ion{C}{2}  &4        &   0&.0264 &   0&.0367 & 15&  \\
 3924&.920& 3921&.88&              &         &   0&.0163 &   0&.0227 & 20&  \\
 3929&.582& 3926&.54& ~\ion{He}{1} &58       &   0&.0875 &   0&.1219 & 10&  \\
 3967&.799& 3964&.73& ~\ion{He}{1} &5        &   0&.6266 &   0&.8605 &  3&  \\
 3970&.540& 3967&.46& [\ion{Ne}{3}]&F1       &   7&.656  &  10&.51   &  1&.2\\
 3973&.150& 3970&.07& ~\ion{H}{1}  &H7       &  11&.29   &  15&.48   &  1&.2\\
 3982&.888& 3979&.78& [\ion{Fe}{2}]&F9       &   0&.0192 &   0&.0263 & 20&  \\
 \mcnd    & 3979&.93& [\ion{Fe}{2}]&F8       &   \mcnd   &   \mcnd   & \mcnd\\
 4007&.490& 4004&.38&              &         &   0&.0183 &   0&.0249 & 20&  \\
 4011&.451& 4008&.3 & [\ion{Fe}{3}]&F4       &   0&.0209 &   0&.0284 & 15&  \\
 4012&.374& 4009&.26& ~\ion{He}{1} &55       &   0&.1346 &   0&.1826 &  7&  \\
 4017&.176& 4013&.99& ~\ion{Ne}{1} &2        &   0&.0231 &   0&.0312 & 15&  \\
 4027&.030& 4023&.98& ~\ion{He}{1} &54       &   0&.0118 &   0&.0160 & 20&  \\
 4029&.331& 4026&.21& ~\ion{He}{1} &18       &   1&.596  &   2&.152  &  2&.5\\
 4069&.167& 4066&.01&              &         &   0&.0168 &   0&.0225 & 20&  \\
 4071&.764& 4068&.60& [\ion{S}{2}] &F1       &   0&.6530 &   0&.8713 &  3&  \\
 4072&.916& 4069&.62& ~\ion{O}{2}  &10       &   0&.0775 &   0&.1035 & 10&  \\
 \mcnd    & 4069&.89& ~\ion{O}{2}  &10       &   \mcnd   &   \mcnd   & \mcnd\\
 4075&.339& 4072&.16& ~\ion{O}{2}  &10       &   0&.0400 &   0&.0534 & 12&  \\
 4079&.513& 4076&.35& [\ion{S}{2}] &F1       &   0&.2221 &   0&.2954 &  5&  \\
 4082&.136& 4078&.84& ~\ion{O}{2}  &10       &   0&.0204 &   0&.0272 & 15&  \\
 4087&.772& 4084&.65& ~\ion{O}{2}  &21       &   0&.0517 &   0&.0687 & 10&  \\
 4092&.151& 4089&.29& ~\ion{O}{2}  &48a      &   0&.0333 &   0&.0442 & 12&  \\
 4100&.448& 4097&.25& ~\ion{O}{2}  &20       &   0&.0129 &   0&.0171 & 20&  \\
 \mcnd    & 4097&.26& ~\ion{O}{2}  &48b      &   \mcnd   &   \mcnd   & \mcnd\\
 \mcnd    & 4097&.33& ~\ion{N}{3}  &1        &   \mcnd   &   \mcnd   & \mcnd\\
 4101&.407& 4098&.24& ~\ion{O}{2}  &46a      &   0&.0069 &   0&.0091 & 30&  \\
 \mcnd    & \mcnd   & ~\ion{H}{1}  &Gal      &   \mcnd   &   \mcnd   & \mcnd\\ 
 4104&.914& 4101&.74& ~\ion{H}{1}  &H6       &  19&.82   &  26&.20   &  1&.0\\
 4110&.192& 4107&.07& ~\ion{O}{2}  &47       &   0&.0074 &   0&.0098 & 25&  \\
 4111&.912& 4108&.75& ~\ion{O}{2}  &48       &   0&.0013 &   0&.0017 & 50&  \\
 4113&.941& 4110&.78& ~\ion{O}{2}  &20       &   0&.0198 &   0&.0261 & 15&  \\
 4115&.477& 4112&.03& ~\ion{O}{2}  &21       &   0&.0109 &   0&.0143 & 20&  \\
 \mcnd    & 4112&.10& ~\ion{Ne}{2} &         &   \mcnd   &   \mcnd   & \mcnd\\
 4121&.528& 4118&.10& ~\ion{Ne}{2} &54       &   0&.0301 &   0&.0396 & 15&  \\
 4122&.378& 4119&.22& ~\ion{O}{2}  &20       &   0&.0124 &   0&.0164 & 20&  \\
 4124&.012& 4120&.84& ~\ion{He}{1} &16       &   0&.1979 &   0&.2600 &  6&  \\
 4127&.960& 4124&.78& [\ion{Fe}{2}]&         &   0&.0228 &   0&.0299 & 15&  \\
 4135&.954& 4132&.80& ~\ion{O}{2}  &19       &   0&.0224 &   0&.0293 & 15&  \\
 4141&.313& 4138&.10&              &         &   0&.0394 &   0&.0516 & 12&  \\
 4145&.090& 4141&.96& ~\ion{O}{2}  &106      &   0&.0074 &   0&.0096 & 25&  \\
 \mcnd    & 4142&.08& ~\ion{O}{2}  &106      &   \mcnd   &   \mcnd   & \mcnd\\
 4145&.684& 4142&.47&              &         &   0&.0015 &   0&.0019 & 50&  \\
 4146&.970& 4143&.76& ~\ion{He}{1} &53       &   0&.2220 &   0&.2899 &  5&  \\
 4148&.235& 4145&.10& ~\ion{S}{2}  &44       &   0&.0015 &   0&.0020 & 50&  \\
 4149&.097& 4145&.90& ~\ion{O}{2}  &106      &   0&.0032 &   0&.0041 & 35&  \\
 \mcnd    & 4146&.09& ~\ion{O}{2}  &106      &   \mcnd   &   \mcnd   & \mcnd\\
 4156&.489& 4153&.30& ~\ion{O}{2}  &19       &   0&.0348 &   0&.0454 & 12&  \\
 4159&.578& 4156&.53& ~\ion{O}{2}  &19       &   0&.0046 &   0&.0060 & 30&  \\
 4160&.679& 4157&.45&              &         &   0&.0279 &   0&.0364 & 15&  \\
 4166&.617& 4163&.26& ~\ion{C}{3}? &21       &   0&.0044 &   0&.0057 & 30&  \\
 4172&.296& 4168&.97& ~\ion{He}{1} &52       &   0&.0504 &   0&.0655 & 10&  \\
 \mcnd    & 4169&.22& ~\ion{O}{2}  &19       &   \mcnd   &   \mcnd   & \mcnd\\
 4188&.693& 4185&.45& ~\ion{O}{2}  &36       &   0&.0114 &   0&.0147 & 20&  \\
 4198&.171& 4194&.92&              &         &   0&.0235 &   0&.0303 & 15&  \\
 4202&.594& 4199&.34&              &         &   0&.0119 &   0&.0153 & 20&  \\
 4206&.608& 4203&.27& ~\ion{Ne}{2} &         &   0&.0095 &   0&.0122 & 25&  \\
 \mcnd    & 4203&.41& ~\ion{Ar}{2} &         &   \mcnd   &   \mcnd   & \mcnd\\
 4207&.793& 4204&.54& ~\ion{Cl}{2} &         &   0&.0141 &   0&.0181 & 20&  \\
 4225&.088& 4221&.80& ~\ion{Cl}{2} &         &   0&.0097 &   0&.0124 & 25&  \\
 4236&.333& 4233&.32& ~\ion{O}{1}  &         &   0&.0223 &   0&.0284 & 15&  \\
 4240&.751& 4237&.47&              &         &   0&.0167 &   0&.0212 & 20&  \\
 4270&.481& 4267&.15& ~\ion{C}{2}  &6        &   0&.0737 &   0&.0927 & 10&  \\
 4275&.167& 4272&.17& ~\ion{Ar}{1} &         &   0&.0298 &   0&.0374 & 15&  \\
 4279&.838& 4276&.28& ~\ion{O}{2}  &67b      &   0&.0227 &   0&.0284 & 15&  \\
 \mcnd    & 4276&.75& ~\ion{O}{2}  &67b      &   \mcnd   &   \mcnd   & \mcnd\\
 \mcnd    & 4276&.83& [\ion{Fe}{2}]&F21      &   \mcnd   &   \mcnd   & \mcnd\\
 4284&.820& 4281&.32& ~\ion{O}{2}  &53b      &   0&.0103 &   0&.0129 & 20&  \\
 4290&.807& 4287&.39& [\ion{Fe}{2}]&F7       &   0&.0365 &   0&.0456 & 12&  \\
 4307&.167& 4303&.61& ~\ion{O}{2}  &65a      &   0&.0464 &   0&.0576 & 10&  \\
 \mcnd    & 4303&.82& ~\ion{O}{2}  &53a      &   \mcnd   &   \mcnd   & \mcnd\\
 4314&.790& 4311&.45&              &         &   0&.0225 &   0&.0279 & 15&  \\
 4319&.382& 4315&.80& ~\ion{O}{2}  &78       &   0&.0123 &   0&.0152 & 20&  \\
 \mcnd    & 4316&.00& ~\ion{Ne}{1} &         &   \mcnd   &   \mcnd   & \mcnd\\
 4320&.635& 4317&.14& ~\ion{O}{2}  &2        &   0&.0259 &   0&.0320 & 15&  \\
 4323&.049& 4319&.63& ~\ion{O}{2}  &2        &   0&.0071 &   0&.0087 & 25&  \\
 4343&.830& 4340&.47& ~\ion{H}{1}  &H5       &  39&.35   &  48&.32   &  0&.7\\
 4349&.003& 4345&.55& ~\ion{O}{2}  &65c      &   0&.0190 &   0&.0232 & 15&  \\
 \mcnd    & 4345&.56& ~\ion{O}{2}  &2        &   \mcnd   &   \mcnd   & \mcnd\\
 4352&.772& 4349&.43& ~\ion{O}{2}  &2        &   0&.0256 &   0&.0314 & 15&  \\
 4355&.080& 4351&.81& [\ion{Fe}{2}]&         &   0&.0331 &   0&.0405 & 12&  \\
 4362&.703& 4359&.34& [\ion{Fe}{2}]&F7       &   0&.0234 &   0&.0285 & 15&  \\
 4366&.593& 4363&.21& [\ion{O}{3}] &F2       &   2&.638  &   3&.209  &  1&.5\\
 4367&.938& 4364&.73& ~\ion{S}{3}  &7        &   0&.0119 &   0&.0145 & 20&  \\
 4370&.172& 4366&.89& ~\ion{O}{2}  &2        &   0&.0202 &   0&.0245 & 15&  \\
 4371&.727& 4368&.25& ~\ion{O}{1}  &5        &   0&.0174 &   0&.0211 & 15&  \\
 4391&.335& 4387&.93& ~\ion{He}{1} &51       &   0&.4573 &   0&.5503 &  4&  \\
 4396&.084& 4392&.68& ~\ion{S}{2}? &         &   0&.0349 &   0&.0419 & 12&  \\
 4400&.994& 4397&.58&              &         &   0&.0153 &   0&.0183 & 20&  \\
 4405&.831& 4402&.60& [\ion{Fe}{2}]&         &   0&.0076 &   0&.0091 & 25&  \\
 4406&.747& 4402&.99& ~\ion{Ne}{1} &         &   0&.0030 &   0&.0036 & 35&  \\
 4417&.287& 4413&.78& [\ion{Fe}{2}]&F6       &   0&.0180 &   0&.0215 & 15&  \\
 4419&.761& 4416&.27& [\ion{Fe}{2}]&F6       &   0&.0196 &   0&.0233 & 15&  \\
 4420&.436& 4416&.97& ~\ion{O}{2}  &5        &   0&.0158 &   0&.0188 & 20&  \\
 4421&.319& 4417&.82& ~\ion{N}{2}  &         &   0&.0026 &   0&.0031 & 40&  \\
 4438&.054& 4434&.68& ~\ion{S}{2}? &         &   0&.0299 &   0&.0353 & 15&  \\
 4441&.010& 4437&.55& ~\ion{He}{1} &50       &   0&.0479 &   0&.0565 & 12&  \\
 4442&.780& 4439&.30& ~\ion{Ne}{2} &65       &   0&.0110 &   0&.0130 & 20&  \\
 \mcnd    & 4439&.46& ~\ion{Ar}{2} &127      &   \mcnd   &   \mcnd   & \mcnd\\
 4455&.703& 4452&.37& ~\ion{O}{2}  &5        &   0&.0166 &   0&.0195 & 20&  \\
 4474&.978& 4471&.50& ~\ion{He}{1} &14       &   3&.709  &   4&.312  &  1&.5\\
 4480&.792& 4477&.47& ~\ion{C}{1}  &         &   0&.0307 &   0&.0356 & 12&  \\
 4485&.561& 4482&.08&              &         &   0&.0091 &   0&.0106 & 25&  \\
 4490&.858& 4487&.46& ~\ion{S}{3}  &2        &   0&.0111 &   0&.0128 & 20&  \\
 4494&.725& 4491&.23& ~\ion{O}{2}  &86a      &   0&.0145 &   0&.0167 & 20&  \\
 4503&.708& 4500&.18& ~\ion{Ne}{1} &         &   0&.0050 &   0&.0058 & 30&  \\
 4513&.082& 4509&.61& [\ion{Fe}{2}]&F6       &   0&.0027 &   0&.0032 & 40&  \\
 4524&.230& 4520&.72&              &         &   0&.0276 &   0&.0314 & 15&  \\
 4529&.114& 4525&.76& ~\ion{Ne}{1} &         &   0&.0160 &   0&.0182 & 20&  \\
 4530&.522& 4527&.01&              &         &   0&.0031 &   0&.0036 & 35&  \\
 4531&.479& 4527&.96& ~\ion{S}{3}  &7        &   0&.0037 &   0&.0043 & 35&  \\
 4533&.024& 4529&.48& ~\ion{Ne}{1} &         &   0&.0051 &   0&.0058 & 30&  \\
 4566&.142& 4562&.60& ~\ion{Mg}{1}]&         &   0&.0638 &   0&.0716 & 10&  \\
 4568&.657& 4565&.12&              &         &   0&.0339 &   0&.0378 & 12&  \\
 4573&.449& 4569&.90&              &         &   0&.0132 &   0&.0148 & 20&  \\
 4574&.667& 4571&.10& ~\ion{Mg}{1}]&1        &   0&.0552 &   0&.0617 & 10&  \\
 4578&.791& 4575&.06& ~\ion{Ne}{1} &         &   0&.0062 &   0&.0069 & 25&  \\
 4594&.630& 4590&.97& ~\ion{O}{2}  &15       &   0&.0080 &   0&.0089 & 25&  \\
 4599&.687& 4595&.96& ~\ion{O}{2}  &15       &   0&.0119 &   0&.0133 & 20&  \\
 \mcnd    & 4596&.18& ~\ion{O}{2}  &15       &   \mcnd   &   \mcnd   & \mcnd\\
 4610&.631& 4607&.06& [\ion{Fe}{3}]&F3       &   0&.0201 &   0&.0221 & 15&  \\
 4613&.803& 4610&.20& ~\ion{O}{2}  &92e      &   0&.0353 &   0&.0388 & 12&  \\
 4618&.754& 4615&.17&              &         &   0&.0120 &   0&.0132 & 20&  \\
 4624&.194& 4620&.61&              &         &   0&.0090 &   0&.0099 & 25&  \\
 4634&.273& 4630&.54& ~\ion{N}{2}  &5        &   0&.0095 &   0&.0104 & 25&  \\
 4642&.425& 4638&.86& ~\ion{O}{2}  &1        &   0&.0459 &   0&.0500 & 10&  \\
 4644&.115& 4640&.64& ~\ion{N}{3}  &2        &   0&.0070 &   0&.0077 & 25&  \\
 4645&.415& 4641&.81& ~\ion{O}{2}  &1        &   0&.0597 &   0&.0648 & 10&  \\
 4652&.724& 4649&.13& ~\ion{O}{2}  &1        &   0&.0494 &   0&.0536 & 10&  \\
 4654&.433& 4650&.84& ~\ion{O}{2}  &1        &   0&.0575 &   0&.0623 & 10&  \\
 4660&.012& 4656&.39& ~\ion{Ne}{1} &         &   0&.0447 &   0&.0483 & 12&  \\
 4661&.750& 4658&.10& [\ion{Fe}{3}]&F3       &   0&.4528 &   0&.4886 &  4&  \\
 4665&.181& 4661&.63& ~\ion{O}{2}  &1        &   0&.0672 &   0&.0725 & 10&  \\
 4670&.532& 4667&.  & [\ion{Fe}{3}]&F3       &   0&.0206 &   0&.0221 & 15&  \\
 4677&.146& 4673&.73& ~\ion{O}{2}  &1        &   0&.0099 &   0&.0107 & 25&  \\
 4679&.909& 4676&.24& ~\ion{O}{2}  &1        &   0&.0196 &   0&.0211 & 15&  \\
 4699&.948& 4696&.35& ~\ion{O}{2}  &1        &   0&.0033 &   0&.0035 & 35&  \\
 4705&.242& 4701&.62& [\ion{Fe}{3}]&F3       &   0&.1177 &   0&.1249 &  7&  \\
 4707&.038& 4703&.18& ~\ion{O}{2}  &         &   0&.0400 &   0&.0425 & 12&  \\
 \mcnd    & 4703&.37& ~\ion{Ar}{2} &         &   \mcnd   &   \mcnd   & \mcnd\\
 4711&.950& 4708&.30&              &         &   0&.0126 &   0&.0134 & 20&  \\
 4713&.747& 4710&.00& ~\ion{O}{2}? &         &   0&.0062 &   0&.0066 & 30&  \\
 4715&.013& 4711&.37& [\ion{Ar}{4}]&F1       &   0&.1236 &   0&.1309 &  7&  \\
 4716&.847& 4713&.17& ~\ion{He}{1} &12       &   0&.4552 &   0&.4811 &  4&  \\
 4737&.637& 4733&.9 & [\ion{Fe}{3}]&F3       &   0&.0385 &   0&.0405 & 12&  \\
 4743&.873& 4740&.17& [\ion{Ar}{4}]&F1       &   0&.0942 &   0&.0986 &  7&  \\
 4755&.099& 4751&.34& ~\ion{O}{2}  &         &   0&.0437 &   0&.0456 & 12&  \\
 4758&.458& 4754&.7 & [\ion{Fe}{3}]&F3       &   0&.0858 &   0&.0892 &  8&  \\
 4760&.206& 4756&.52&              &         &   0&.0225 &   0&.0234 & 15&  \\
 4764&.765& 4761&.07&              &         &   0&.0035 &   0&.0036 & 35&  \\
 4765&.880& 4762&.31& ~\ion{C}{1}  &6        &   0&.0226 &   0&.0235 & 15&  \\
 4773&.174& 4769&.4 & [\ion{Fe}{3}]&F3       &   0&.0526 &   0&.0545 & 10&  \\
 4781&.444& 4777&.7 & [\ion{Fe}{3}]&F3       &   0&.0152 &   0&.0157 & 20&  \\
 4804&.118& 4800&.11& ~\ion{Ne}{1}?&         &   0&.0455 &   0&.0466 & 12&  \\
 4809&.235& 4805&.51&              &         &   0&.0229 &   0&.0234 & 15&  \\
 4818&.314& 4814&.53& [\ion{Fe}{2}]&F20      &   0&.0087 &   0&.0089 & 25&  \\
 4854&.175& 4850&.70& [\ion{Fe}{2}]&F20      &   0&.0553 &   0&.0555 & 10&  \\
 4859&.299& 4855&.53&              &         &   0&.0259 &   0&.0260 & 15&  \\
 4861&.063& \mcnd   & ~\ion{H}{1}  &Gal      &   0&.0061 &   \mcnd   & 35&  \\
 4865&.092& 4861&.33& ~\ion{H}{1}  &H4       & 100&.00   & 100&.00   &  0&.7\\
 4874&.303& 4870&.53&              &         &   0&.0061 &   0&.0061 & 30&  \\
 4882&.950& 4879&.17&              &         &   0&.0124 &   0&.0123 & 20&  \\
 4884&.826& 4881&.11& [\ion{Fe}{3}]&F2       &   0&.1525 &   0&.1512 &  7&  \\
 4905&.195& 4901&.28& ~\ion{S}{2}  &         &   0&.0375 &   0&.0369 & 12&  \\
 4906&.485& 4902&.65& ~\ion{S}{3}  &         &   0&.0168 &   0&.0165 & 20&  \\
 4925&.757& 4921&.93& ~\ion{He}{1} &48       &   1&.157  &   1&.131  &  3&  \\
 4928&.381& 4924&.53& ~\ion{O}{2}  &28       &   0&.0349 &   0&.0341 & 12&  \\
 \mcnd    & 4925&.  & [\ion{Fe}{3}]&F2       &   \mcnd   &   \mcnd   & \mcnd\\
 4934&.994& 4931&.8 & [\ion{O}{3}] &F1       &   0&.0632 &   0&.0615 & 10&  \\
 \mcnd    & 4931&.  & [\ion{Fe}{3}]&F1       &   \mcnd   &   \mcnd   & \mcnd\\
 4937&.616& 4933&.79&              &         &   0&.0047 &   0&.0045 & 30&  \\
 4962&.783& 4958&.91& [\ion{O}{3}] &F1       & 176&.1    & 169&.9    &  0&.7\\
 4989&.675& 4985&.9 & [\ion{Fe}{3}]&F2       &   0&.2329 &   0&.2224 &  5&  \\
 4991&.102& 4987&.2 & [\ion{Fe}{3}]&F2       &   0&.0451 &   0&.0431 & 12&  \\
 5010&.752& 5006&.84& [\ion{O}{3}] &F1       & 537&.3    & 508&.9    &  0&.7\\
 5015&.177& 5011&.3 & [\ion{Fe}{3}]&F1       &   0&.0522 &   0&.0493 & 12&  \\
 5019&.581& 5015&.68& ~\ion{He}{1} &4        &   2&.335  &   2&.204  &  2&.0\\
 5036&.267& 5032&.40& [\ion{Fe}{4}]&1        &   0&.0179 &   0&.0168 & 20&  \\
 \mcnd    & 5032&.43& ~\ion{S}{2}  &7        &   \mcnd   &   \mcnd   & \mcnd\\
 5039&.706& 5035&.71& [\ion{Fe}{2}]&         &   0&.0110 &   0&.0103 & 25&  \\
 5045&.037& 5041&.03& ~\ion{S}{3}  &5        &   0&.1390 &   0&.1301 &  7&  \\
 5060&.002& 5055&.98& ~\ion{S}{3}  &5        &   0&.0955 &   0&.0888 &  7&  \\
 \mcnd    & 5056&.31& ~\ion{S}{3}  &5        &   \mcnd   &   \mcnd   & \mcnd\\
 5162&.842& 5158&.78& [\ion{Fe}{2}]&F19      &   0&.0341 &   0&.0306 & 15&  \\
 5195&.753& 5191&.82& [\ion{Ar}{3}]&F3       &   0&.0749 &   0&.0663 & 10&  \\
 5202&.041& 5197&.90& [\ion{N}{1}] &F1       &   0&.1457 &   0&.1288 &  7&  \\
 5204&.403& 5200&.26& [\ion{N}{1}] &F1       &   0&.0784 &   0&.0693 & 10&  \\
 5265&.744& 5261&.62& [\ion{Fe}{2}]&F19      &   0&.0229 &   0&.0198 & 15&  \\
 5274&.617& 5270&.4 & [\ion{Fe}{3}]&F1       &   0&.2836 &   0&.2441 &  5&  \\
 5312&.844& 5308&.73&              &         &   0&.0040 &   0&.0034 & 35&  \\
 5328&.697& 5324&.61&              &         &   0&.0092 &   0&.0077 & 25&  \\
 5329&.330& 5325&.20&              &         &   0&.0153 &   0&.0129 & 20&  \\
 5330&.349& 5326&.40& ~\ion{Ne}{1}?&         &   0&.0230 &   0&.0194 & 15&  \\
 5416&.362& 5412&.  & [\ion{Fe}{3}]&F1       &   0&.0237 &   0&.0194 & 15&  \\
 5521&.982& 5517&.71& [\ion{Cl}{3}]&F1       &   0&.5690 &   0&.4487 &  4&  \\
 5542&.146& 5537&.88& [\ion{Cl}{3}]&F1       &   0&.4231 &   0&.3319 &  4&  \\
 5644&.997& 5640&.32& ~\ion{S}{2}  &11       &   0&.0100 &   0&.0075 & 25&  \\
 \mcnd    & 5640&.55& ~\ion{C}{2}  &15       &   \mcnd   &   \mcnd   & \mcnd\\
 5683&.942& 5679&.56& ~\ion{N}{2}  &3        &   0&.0118 &   0&.0087 & 25&  \\
 5700&.807& 5696&.47& ~\ion{Al}{3}?&2        &   0&.0036 &   0&.0027 & 40&  \\
 5759&.097& 5754&.64& [\ion{N}{2}] &F3       &   0&.2595 &   0&.1871 &  6&  \\
 5875&.069& 5870&.44& ~\ion{Ar}{2}?&         &   0&.0049 &   0&.0034 & 35&  \\
 5880&.229& 5875&.67& ~\ion{He}{1} &11       &  17&.48   &  12&.12   &  1&.0\\
 5952&.196& 5947&.58&              &         &   0&.0044 &   0&.0030 & 35&  \\
 5962&.239& 5957&.56& ~\ion{S}{3}  &4        &   0&.0272 &   0&.0183 & 15&  \\
 5963&.266& 5958&.58& ~\ion{O}{1}  &23       &   0&.0249 &   0&.0167 & 20&  \\
 5983&.612& 5978&.93& ~\ion{S}{3}  &4        &   0&.0580 &   0&.0388 & 12&  \\
 5992&.109& 5987&.46&              &         &   0&.0202 &   0&.0134 & 20&  \\
 6051&.155& 6046&.40& ~\ion{O}{1}  &22       &   0&.0219 &   0&.0143 & 20&  \\
 6155&.547& 6150&.90& ~\ion{N}{2}  &36       &   0&.0223 &   0&.0142 & 20&  \\
 6156&.319& 6151&.43& ~\ion{C}{2}  &16.04    &   0&.0362 &   0&.0229 & 15&  \\
 6210&.265& 6205&.56& ~\ion{C}{3}? &198\TA   &   0&.0081 &   0&.0050 & 30&  \\
 \mcnd    & 6205&.77& ~\ion{Ne}{1} &         &   \mcnd   &   \mcnd   & \mcnd\\
 6305&.246& 6300&.30& [\ion{O}{1}] &F1       &   1&.522  &   0&.9266 &  2&.5\\
 6317&.003& 6312&.10& [\ion{S}{3}] &F3       &   2&.976  &   1&.805  &  2&.0\\
 6352&.073& 6347&.09& ~\ion{S}{3}  &2        &   0&.1274 &   0&.0766 & 10&  \\
 6368&.773& 6363&.78& [\ion{O}{1}] &F1       &   0&.5426 &   0&.3248 &  4&  \\
 6376&.355& 6371&.36& ~\ion{S}{3}  &2        &   0&.0978 &   0&.0585 & 10&  \\
 6553&.191& 6548&.03& [\ion{N}{2}] &F1       &   6&.601  &   3&.763  &  1&.5\\
 6555&.778& 6550&.60& ~\ion{O}{2}  &         &   0&.0196 &   0&.0112 & 20&  \\
 6561&.234& 6556&.06& ~\ion{O}{2}  &113      &   0&.0064 &   0&.0036 & 35&  \\
 6562&.322& \mcnd   & ~\ion{H}{1}  &Gal      &   0&.0502 &   \mcnd   & 20&  \\
 6567&.913& 6562&.82& ~\ion{H}{1}  &H3       & 502&.3    & 285&.3    &  0&.7\\
 6583&.170& 6578&.05& ~\ion{C}{2}  &2        &   0&.1160 &   0&.0656 & 10&  \\
 6588&.591& 6583&.41& [\ion{N}{2}] &F1       &  20&.31   &  11&.47   &  1&.0\\
 6596&.170& 6591&.06&              &         &   0&.0255 &   0&.0144 & 20&  \\
 6627&.170& 6622&.05& ~\ion{C}{2}  &         &   0&.0142 &   0&.0079 & 25&  \\
 6683&.368& 6678&.15& ~\ion{He}{1} &46       &   6&.131  &   3&.385  &  1&.5\\
 6721&.716& 6716&.47& [\ion{S}{2}] &F2       &  13&.02   &   7&.122  &  1&.2\\
 6736&.115& 6730&.85& [\ion{S}{2}] &F2       &  12&.04   &   6&.558  &  1&.2\\
 6739&.390& 6734&.00& ~\ion{C}{2}? &132\TA   &   0&.0419 &   0&.0228 & 15&  \\
 \mcnd    & 6733&.9 & [\ion{Cr}{4}]&         &   \mcnd   &   \mcnd   & \mcnd\\
 6752&.837& 6747&.5 & [\ion{Cr}{4}]&         &   0&.0193 &   0&.0105 & 20&  \\
 7007&.671& 7002&.13& ~\ion{O}{1}  &21       &   0&.0571 &   0&.0292 & 15&  \\
 7070&.772& 7065&.25& ~\ion{He}{1} &10       &   5&.608  &   2&.833  &  2&.0\\
 7116&.377& 7110&.9 & [\ion{Cr}{4}]&         &   0&.0241 &   0&.0120 & 20&  \\
 7141&.357& 7135&.80& [\ion{Ar}{3}]&F1       &  24&.09   &  11&.98   &  1&.0\\
 7160&.811& 7155&.16& [\ion{Fe}{2}]&F14      &   0&.0418 &   0&.0207 & 15&  \\
 7287&.045& 7281&.35& ~\ion{He}{1} &45       &   0&.9357 &   0&.4518 &  4&  \\
 7324&.771& 7318&.92& [\ion{O}{2}] &F2       &   1&.062  &   0&.5087 &  4&  \\
 7325&.847& 7319&.99& [\ion{O}{2}] &F2       &   3&.224  &   1&.544  &  2&.5\\
 7335&.435& 7329&.67& [\ion{O}{2}] &F2       &   1&.761  &   0&.8420 &  3&  \\
 7336&.496& 7330&.73& [\ion{O}{2}] &F2       &   1&.744  &   0&.8339 &  3&  \\
 7383&.776& 7377&.83& [\ion{N}{3}] &F2       &   0&.0183 &   0&.0087 & 25&  \\
 7448&.232& 7442&.3 & ~\ion{N}{1}  &3        &   0&.0393 &   0&.0183 & 20&  \\
 7458&.185& 7452&.54& [\ion{Fe}{2}]&F14      &   0&.0175 &   0&.0081 & 25&  \\
 7474&.221& 7468&.29& ~\ion{N}{1}  &3        &   0&.0534 &   0&.0248 & 15&  \\
 7487&.597& 7481&.79&              &         &   0&.0060 &   0&.0027 & 40&  \\
 7505&.736& 7499&.82& ~\ion{He}{1} &1/8\TB   &   0&.0696 &   0&.0322 & 15&  \\
 7536&.374& 7530&.54& [\ion{Cl}{4}]&F1       &   0&.0339 &   0&.0155 & 20&  \\
 7589&.606& 7583&.72&              &         &   0&.0032 &   0&.0014 & 50&  \\
 7692&.987& 7687&.90& [\ion{Fe}{2}]&F1       &   0&.0042 &   0&.0018 & 50&  \\
 7757&.184& 7751&.12& [\ion{Ar}{3}]&F1       &   6&.296  &   2&.774  &  2&.0\\
 7768&.666& 7762&.24& ~\ion{N}{2}  &153\TA   &   0&.0091 &   0&.0040 & 35&  \\
 7778&.189& 7771&.96& ~\ion{O}{1}  &1        &   0&.0115 &   0&.0051 & 30&  \\
 7781&.641& 7775&.40& ~\ion{O}{1}  &1        &   0&.0066 &   0&.0032 & 40&  \\
 7797&.239& 7790&.98& ~\ion{Mg}{2}?&         &   0&.0064 &   0&.0028 & 40&  \\
 7822&.265& 7816&.16& ~\ion{He}{1} &69       &   0&.1079 &   0&.0469 & 12&  \\
 7877&.791& 7871&.69&              &         &   0&.0220 &   0&.0094 & 25&  \\
 7932&.914& 7926&.90& [\ion{Fe}{2}]&F1       &   0&.0123 &   0&.0052 & 35&  \\
 7951&.788& 7945&.62&              &         &   0&.0144 &   0&.0061 & 30&  \\
 8051&.953& 8045&.63& [\ion{Cl}{4}]&F1       &   0&.0816 &   0&.0339 & 15&  \\
 8256&.689& 8249&.97& ~\ion{H}{1}  &P40      &   0&.0441 &   0&.0176 & 20&  \\
 8258&.575& 8252&.40& ~\ion{H}{1}  &P39      &   0&.0457 &   0&.0183 & 20&  \\
 8261&.467& 8255&.02& ~\ion{H}{1}  &P38      &   0&.0883 &   0&.0353 & 15&  \\
 8264&.348& 8257&.86& ~\ion{H}{1}  &P37      &   0&.0965 &   0&.0386 & 15&  \\
 8267&.420& 8260&.94& ~\ion{H}{1}  &P36      &   0&.1133 &   0&.0453 & 15&  \\
 8270&.721& 8264&.29& ~\ion{H}{1}  &P35      &   0&.1209 &   0&.0483 & 12&  \\
 8274&.640& 8267&.94& ~\ion{H}{1}  &P34      &   0&.1639 &   0&.0654 & 12&  \\
 8282&.691& 8276&.31& ~\ion{H}{1}  &P32\TC   &   0&.2236 &   0&.0891 & 10&  \\
 8292&.913& 8286&.43& ~\ion{H}{1}  &P30      &   0&.1953 &   0&.0776 & 12&  \\
 8298&.752& 8292&.31& ~\ion{H}{1}  &P29\TC   &   0&.2074 &   0&.0824 & 10&  \\
 8305&.359& 8298&.84& ~\ion{H}{1}  &P28      &   0&.2244 &   0&.0890 & 10&  \\
 8312&.627& 8306&.22& ~\ion{H}{1}  &P27      &   0&.2388 &   0&.0947 & 10&  \\
 8329&.998& 8323&.43& ~\ion{H}{1}  &P25      &   0&.3003 &   0&.1186 & 10&  \\
 8340&.307& 8333&.78& ~\ion{H}{1}  &P24      &   0&.3496 &   0&.1379 &  7&  \\
 8365&.521& 8359&.01& ~\ion{H}{1}  &P22      &   0&.4894 &   0&.1922 &  7&  \\
 8368&.257& 8361&.77& ~\ion{He}{1} &68       &   0&.2055 &   0&.0805 & 10&  \\
 8381&.017& 8374&.48& ~\ion{H}{1}  &P21      &   0&.5620 &   0&.2198 &  7&  \\
 8419&.911& 8413&.32& ~\ion{H}{1}  &P19      &   0&.7086 &   0&.2749 &  6&  \\
 8444&.568& 8437&.96& ~\ion{H}{1}  &P18      &   0&.9159 &   0&.3538 &  5&  \\
 8451&.188& 8444&.34& ~\ion{N}{3}  &267\TA   &   0&.0777 &   0&.0299 & 15&  \\
 \mcnd    & 8444&.34& ~\ion{He}{1} &4/11\TB  &   \mcnd   &   \mcnd   & \mcnd\\
 8453&.148& 8446&.48& ~\ion{O}{1}  &4        &   0&.6543 &   0&.2522 &  6&  \\
 8457&.771& 8451&.0 & ~\ion{He}{1}?&6/17\TB  &   0&.0275 &   0&.0106 & 30&  \\
 \mcnd    & 8451&.55& ~\ion{S}{1}  &14       &   \mcnd   &   \mcnd   & \mcnd\\
 8460&.473& 8453&.50& ~\ion{He}{1}?&7/17\TB  &   0&.0210 &   0&.0080 & 30&  \\
 8473&.892& 8467&.26& ~\ion{H}{1}  &P17      &   1&.016  &   0&.3898 &  5&  \\
 8487&.455& 8480&.9 & [\ion{Cl}{3}]&F3       &   0&.0173 &   0&.0066 & 35&  \\
 8506&.601& 8500&.0 & [\ion{Cl}{3}]&F3       &   0&.0284 &   0&.0108 & 30&  \\
 8509&.160& 8502&.49& ~\ion{H}{1}  &P16      &   1&.191  &   0&.4544 &  5&  \\
 8520&.115& 8513&.51&              &         &   0&.0172 &   0&.0065 & 35&  \\
 8524&.816& 8518&.04& ~\ion{He}{1} &2/8\TB   &   0&.0295 &   0&.0112 & 30&  \\
 8535&.618& 8528&.99& ~\ion{He}{1} &10/17\TB &   0&.0301 &   0&.0114 & 30&  \\
 \mcnd    & 8528&.99& ~\ion{He}{1} &6/15\TB  &   \mcnd   &   \mcnd   & \mcnd\\
 8671&.814& 8665&.02& ~\ion{H}{1}  &P13      &   2&.100  &   0&.7746 &  4&  \\
 8718&.600& 8711&.7 & ~\ion{N}{1}  &1        &   0&.0315 &   0&.0115 & 30&  \\
 8757&.426& 8750&.48& ~\ion{H}{1}  &P12      &   2&.916  &   1&.058  &  4&  \\
 8783&.615& 8776&.77& ~\ion{He}{1} &4/9\TB   &   0&.1020 &   0&.0368 & 15&  \\
 8852&.359& 8845&.38& ~\ion{He}{1} &6/11\TB  &   0&.1412 &   0&.0503 & 15&  \\
 8855&.169& 8848&.05& ~\ion{He}{1} &7/11\TB  &   0&.0638 &   0&.0227 & 20&  \\
 8869&.742& 8862&.79& ~\ion{H}{1}  &P11      &   3&.711  &   1&.317  &  3&  \\
 9004&.146& 8996&.99& ~\ion{He}{1} &6/10\TB  &   0&.1389 &   0&.0480 & 15&  \\
 9070&.199& 9063&.27& ~\ion{He}{1} &4/8\TB   &   0&.2054 &   0&.0703 & 15&  \\
 9076&.056& 9068&.9 & [\ion{S}{3}] &F1       &  83&.35   &  28&.48   &  1&.2\\
 9131&.431& 9123&.6 & [\ion{Cl}{2}]&F1       &   0&.0473 &   0&.0160 & 50&  \\
 9217&.655& 9210&.28& ~\ion{He}{1} &83       &   0&.2408 &   0&.0805 & 15&  \\
 9220&.429& 9213&.24& ~\ion{He}{1} &7/9\TB   &   0&.0752 &   0&.0251 & 25&  \\
 9236&.236& 9229&.02& ~\ion{H}{1}  &P9       &   7&.775  &   2&.594  &  3&  \\
 9251&.488& 9244&.3 & ~\ion{Mg}{2} &1        &   0&.0287 &   0&.0095 & 35&  \\
 9340&.146& 9332&.91&              &         &   0&.0863 &   0&.0283 & 25&  \\
 9350&.875& 9343&.63&              &         &   0&.1463 &   0&.0479 & 20&  \\
 9471&.120& 9463&.57& ~\ion{He}{1} &67       &   0&.5599 &   0&.1800 & 10&  \\
 9533&.801& 9526&.17& ~\ion{He}{1} &6/8\TB   &   0&.5500 &   0&.1753 & 10&  \\
 9538&.536& 9531&.10& [\ion{S}{3}] &F1       & 204&.8    &  65&.28   &  1&.2\\
 9684&.909& 9677&.38& ~\ion{O}{1}  &         &   0&.0299 &   0&.0093 & 40&  \\
 9704&.129& 9696&.51& ~\ion{C}{3}? &         &   0&.0190 &   0&.0059 & 50&  \\
 9858&.221& 9850&.24& [\ion{C}{1}] &F1       &   0&.0686 &   0&.0209 & 30&  \\
10020&.60 &10012&.8 & ~\ion{O}{2}  &         &   0&.1113 &   0&.0332 & 30&  \\
10035&.67 &10027&.7 & ~\ion{He}{1} &6/7\TB   &   0&.5822 &   0&.1732 & 12&  \\
10039&.08 &10031&.2 & ~\ion{He}{1} &7/7\TB   &   0&.1832 &   0&.0545 & 20&  \\
10057&.29 &10049&.4 & ~\ion{H}{1}  &P7       &  14&.70   &   4&.365  &  3&  \\
10145&.99 &10138&.5 & ~\ion{He}{1} &10/7\TB  &   0&.1593 &   0&.0468 & 25&  \\
10166&.97 &10159&.1 &              &         &   0&.1034 &   0&.0303 & 30&  \\
10291&.71 &10284&.2 & ~\ion{C}{1}? &         &   0&.2370 &   0&.0684 & 20&  \\
10319&.35 &10311&.3 & ~\ion{He}{1} &4/6\TB   &   0&.3487 &   0&.1003 & 20&  \\
10328&.51 &10320&.4 & [\ion{S}{2}] &F3       &   0&.6402 &   0&.1839 & 15&  \\
10339&.32 &10331&.3 &              &         &   0&.2068 &   0&.0594 & 25&  \\
10344&.41 &10336&.3 & [\ion{S}{2}] &F3       &   0&.2302 &   0&.0661 & 25&  \\
\enddata
\tablenotetext{a}{See \citet{wie96}.}
\tablenotetext{b}{See \citet{peq88}.}
\tablenotetext{c}{Contaminated by a telluric line in emission.}
\end{deluxetable} 

\clearpage

\begin{deluxetable}{l@{\hspace{12pt}}r@{}l@{\hspace{24pt}}l@{\hspace{12pt}}r@{}l}
\tablecaption{Temperatures and Densities
\label{ttemden}}
\tablewidth{0pt}
\tablehead{
\colhead{Lines}  &
\multicolumn{2}{c}{$T_e$ (K)} &
\colhead{Lines} &
\multicolumn{2}{c}{$N_e$(cm$^{-3}$)}}
\startdata
{[\ion{S}{2}]}  & $ 9190$&$\pm 190$         & [\ion{N}{1}]  & $2600$&$^{+1700}_{-1000}$ \\
{[\ion{N}{2}]}  & $10800$&$\pm 300$         & [\ion{S}{2}]  & $ 415$&$\pm 35$           \\
{[\ion{O}{2}]}  & $11050$&$\pm 150$         & [\ion{O}{2}]  & $ 279$&$\pm 16$           \\
{[\ion{S}{3}]}  & $10850$&$\pm 125$         & [\ion{Fe}{3}] & $ 295$&$\pm 30$           \\
{[\ion{Ar}{3}]} & $ 9140$&$^{+310}_{-260}$  & [\ion{Cl}{3}] & $ 270$&$^{+250}_{-230}$   \\
{[\ion{Cl}{3}]} & $10200$&$^{+3000}_{-1500}$& [\ion{Ar}{4}] & $ 700$&$\pm 1200$         \\
{[\ion{O}{3}]}  & $ 9950$&$\pm 60$          \\
$T$(Bac)        & $ 9220$&$\pm 350$         \\
\enddata
\end{deluxetable}

\clearpage

\begin{deluxetable}{l@{\hspace{24pt}}cr@{}lr@{}l}
\tablecaption{Emission line ratios for transitions among the $3d^6$ levels of
\ion{Fe}{3}\tablenotemark{a}
\label{tFepred}}
\tablewidth{0pt}
\tablehead{
\colhead{$\lambda_0$} &
\colhead{Transition} &
\multicolumn{2}{c}{Observed} &
\multicolumn{2}{c}{Predicted}}
\startdata
4607  &  $^3$F$_3$-$^5$D$_4$    &   4.54&$\pm$0.68    &    5.&23 \\
4658  &  $^3$F$_4$-$^5$D$_4$    & 100.00&$\pm$4.00    &  100.&00 \\
4702  &  $^3$F$_3$-$^5$D$_3$    &  25.60&$\pm$1.79    &   28.&10 \\
4734  &  $^3$F$_2$-$^5$D$_2$    &   8.30&$\pm$1.00    &    7.&71 \\ 
4755  &  $^3$F$_4$-$^5$D$_3$    &  18.30&$\pm$1.46    &   18.&80 \\ 
4769  &  $^3$F$_3$-$^5$D$_2$    &  11.20&$\pm$1.12    &    9.&70 \\ 
4778  &  $^3$F$_2$-$^5$D$_1$    &   3.22&$\pm$0.64    &    3.&73 \\
4881  &  $^3$H$_4$-$^5$D$_4$    &  31.00&$\pm$2.17    &   42.&20 \\ 
4986  &  $^3$H$_6$-$^5$D$_4$    &  45.50&$\pm$2.28    &   43.&80 \\
4987  &  $^3$H$_4$-$^5$D$_3$    &   8.82&$\pm$1.06    &    8.&25 \\ 
5011  &  $^3$P$_1$-$^5$D$_2$    &  10.10&$\pm$1.21    &   11.&10 \\
5270  &  $^3$P$_2$-$^5$D$_3$    &  50.00&$\pm$2.50    &   51.&40 \\
5412  &  $^3$P$_2$-$^5$D$_1$    &   3.98&$\pm$0.60    &    4.&84 \\
\enddata
\tablenotetext{a}{Line intensity ratios 100~$\times~I(\lambda_{nm})$/$I$(4658), 
for $T_e$ = 10,000 K and $N_e$ = 316 cm$^{-3}$.}
\end{deluxetable}

\clearpage

\begin{deluxetable}{l@{\hspace{24pt}}c}
\tablecaption{$t^2$ Parameter 
\label{tt2}}
\tablewidth{0pt}
\tablehead{
\colhead{Method} & \colhead{$t^2$}} 
\startdata
$T$(Bac)/$T$(FL)   & $0.022 \pm 0.007$ \\
O$^+$(R/C)         & $0.075 \pm 0.040$ \\
O$^{++}$(R/C)      & $0.038 \pm 0.005$ \\
C$^{++}$(R/C)      & $0.056 \pm 0.040$ \\
\hline
Adopted            & $0.033 \pm 0.005$ \\
\enddata
\end{deluxetable}

\clearpage

\begin{deluxetable}{l@{\hspace{24pt}}r@{}lr@{}l}
\tablecaption{He$^+$ Ionic Abundance 
\label{thelium}}
\tablewidth{0pt}
\tablehead{
\colhead{} &
\multicolumn{4}{c}{He$^+$/H$^+$\tablenotemark{a}}\\
\colhead{} &
\multicolumn{4}{c}{$t^2=0.033$}\\
\cline{2-5}
\colhead{Line} &
\multicolumn{2}{c}{$\tau_{3889}=4.4$} &
\multicolumn{2}{c}{$\tau_{3889}=10.5$}} 
\startdata
3188    & 7167&$\pm303$\tablenotemark{b} & 8825&$\pm368$ \\ 
3614    & 7640&$\pm507$                  & 7640&$\pm507$ \\ 
3819    & 8730&$\pm294$                  & 8730&$\pm294$ \\ 
3889    & 5731&$\pm357$\tablenotemark{b} & 7516&$\pm460$ \\ 
4388    & 8892&$\pm349$                  & 8892&$\pm349$ \\ 
4471    & 8459&$\pm128$                  & 8409&$\pm127$ \\ 
4713    & 8788&$\pm346$                  & 7906&$\pm316$\tablenotemark{c} \\ 
4922    & 8392&$\pm215$                  & 8392&$\pm215$ \\ 
5876    & 8441&$\pm 86$                  & 8291&$\pm 84$ \\ 
6678    & 8412&$\pm133$                  & 8412&$\pm133$ \\ 
7065    & 8459&$\pm147$                  & 5389&$\pm 98$\tablenotemark{c} \\ 
\hline
Adopted & 8470&$\pm 68$\tablenotemark{d} \\ 
\\
\enddata
\tablenotetext{a}{Given in units of $10^{-5}$.}
\tablenotetext{b}{These lines require a higher $\tau_{3889}$ value to be consistent
with the helium abundance and are not included in the adopted average, see
text.}
\tablenotetext{c}{These lines require a lower $\tau_{3889}$ value to be consistent
with the helium abundance, see
text.}
\tablenotetext{d}{Includes the effects of the uncertainty in
$t^2=0.033\pm0.005$.}
\end{deluxetable}

\clearpage

\begin{deluxetable}{l@{\hspace{48pt}}c}
\tablewidth{0pt}
\tablecaption{
C and O ionic abundances from recombination lines\tablenotemark{a}
\label{trl}}
\tablehead{
\colhead{Ion} & \colhead{30 Dor} }
\startdata
C$^{++}$/H$^+$ & 7.96$\pm$0.04  \\
O$^+$/H$^+$    & 7.81$\pm$0.12  \\
O$^{++}$/H$^+$ & 8.46$\pm$0.02  \\
\enddata
\tablenotetext{a}{In units of 12+log(X$^m$/H$^+$).}
\end{deluxetable}

\clearpage

\begin{deluxetable}{l@{\hspace{24pt}}cc}
\tablecaption{
Ionic abundances from collisionally excited lines\tablenotemark{a}
\label{tcl}}
\tablewidth{0pt}
\tablehead{
\colhead{Ion} & \colhead{$t^2$ = 0.00} & \colhead{$t^2$ = 0.033} }
\startdata
C$^{++}$   & 7.71$\pm$0.20\tablenotemark{b}& 7.83$\pm$0.20 \\
N$^+$      & 6.27$\pm$0.02                 & 6.39$\pm$0.02 \\
O$^+$      & 7.55$\pm$0.03                 & 7.69$\pm$0.03 \\
O$^{++}$   & 8.25$\pm$0.01                 & 8.43$\pm$0.01 \\ 
Ne$^{++}$  & 7.57$\pm$0.01                 & 7.77$\pm$0.01 \\ 
S$^+$      & 5.43$\pm$0.02                 & 5.55$\pm$0.02 \\ 
S$^{++}$   & 6.61$\pm$0.01                 & 6.74$\pm$0.01 \\
Cl$^+$     & 3.69$\pm$0.20                 & 3.79$\pm$0.20 \\ 
Cl$^{++}$  & 4.69$\pm$0.02                 & 4.86$\pm$0.02 \\
Cl$^{+3}$  & 3.47$\pm$0.05                 & 3.61$\pm$0.05 \\ 
Ar$^{++}$  & 6.03$\pm$0.01                 & 6.18$\pm$0.01 \\
Ar$^{+3}$  & 4.31$\pm$0.03                 & 4.50$\pm$0.03 \\
Fe$^{++}$  & 5.35$\pm$0.03                 & 5.48$\pm$0.03 \\
\enddata
\tablenotetext{a}{In units of 12+log(X$^m$/H$^+$).}
\tablenotetext{b}{\citet{duf82,gar95}.}
\end{deluxetable}

\clearpage

\begin{deluxetable}{l@{\hspace{24pt}}r@{$\pm$}lr@{$\pm$}l}
\tablecaption{30~Doradus Total Abundances\tablenotemark{a}
\label{ttotab}}
\tablewidth{0pt}
\tablehead{
\colhead{Element}  & 
\multicolumn{2}{c}{$t^2 = 0.00$}  & 
\multicolumn{2}{c}{$t^2$ = 0.033}} 
\startdata
He\tablenotemark{b}& 10.938 & 0.003   & 10.928 & 0.003   \\
C                  & 7.90   & 0.20    & 8.02   & 0.20    \\
C\tablenotemark{b} & 8.15   & 0.05    & 8.15   & 0.05    \\
N                  & 7.05   & 0.08    & 7.21   & 0.08    \\
O                  & 8.41   & 0.02    & 8.58   & 0.02    \\
O\tablenotemark{b} & 8.62   & 0.05    & 8.62   & 0.06    \\
Ne                 & 7.65   & 0.06    & 7.83   & 0.06    \\
S                  & 6.84   & 0.10    & 6.99   & 0.10    \\
Cl                 & 4.75   & 0.12    & 4.82   & 0.12    \\
Ar                 & 6.09   & 0.10    & 6.26   & 0.10    \\
Fe                 & 6.25   & 0.20    & 6.39   & 0.20    \\
\enddata
\tablenotetext{a}{In units of 12 + Log $N$(X)/$N$(H). Gaseous content with the
exception of O and C where 0.08 dex and 0.10 dex have been added respectively
to include the fraction of these elements trapped in dust grains
\citep{est98}.}
\tablenotetext{b}{Values derived from recombination lines. All the other 
values are based on collisionally excited lines.}
\end{deluxetable}

\clearpage

\begin{deluxetable}{lr@{$\pm$}lr@{$\pm$}lr@{$\pm$}lr@{$\pm$}lr@{$\pm$}l}
\tablecaption{Solar and \ion{H}{2} Regions Total Abundances\tablenotemark{a}
\label{tta}}
\tablewidth{0pt}
\tablehead{
\colhead{Element}  &
\multicolumn{2}{c}{30~Doradus\tablenotemark{b}} &
\multicolumn{2}{c}{NGC~346\tablenotemark{c}} & 
\multicolumn{2}{c}{Orion \tablenotemark{d}} &
\multicolumn{2}{c}{M17\tablenotemark{e}} &
\multicolumn{2}{c}{Sun\tablenotemark{f}}}
\startdata
12 + log O/H         & 8.59    & 0.06    & 8.15    & 0.06    & 8.72    & 0.06    & 8.87   & 0.06   & 8.71   & 0.05   \\
log C/O              & -0.45   & 0.05    & -0.87   & 0.08    & -0.19   & 0.08    & -0.11  & 0.08   & -0.21  & 0.10   \\
log N/O              & -1.38   & 0.08    & -1.64   & 0.10    & -0.93   & 0.10    & -0.97  & 0.12   & -0.78  & 0.12   \\
log Ne/O             & -0.76   & 0.06    & -0.85   & 0.06    & -0.82   & 0.12    & -0.85  & 0.12   & -0.71  & 0.09   \\
log S/O              & -1.60   & 0.10    & -1.56   & 0.12    & -1.54   & 0.12    & -1.56  & 0.12   & -1.51  & 0.08   \\
log Cl/O             & -3.67   & 0.12    & \mcnd             & -3.29   & 0.13    & -3.45  & 0.14   & -3.43  & 0.08   \\
log Ar/O             & -2.33   & 0.10    & -2.33   & 0.10    & -2.23   & 0.21    & -2.27  & 0.10   & -2.31  & 0.08   \\
log Fe/O             & -2.20   & 0.20    & -1.99   & 0.15    & -2.61   & 0.20    & -2.18  & 0.16   & -1.21  & 0.06   \\
$X$\tablenotemark{g} & 0.7395  & 0.0020  & 0.7563  & 0.0018  & 0.710   & 0.007   & 0.7124 & 0.0040 & 0.7142 & 0.0100 \\
$Y$\tablenotemark{g} & 0.2506  & 0.0017  & 0.2405  & 0.0018  & 0.276   & 0.007   & 0.2677 & 0.0025 & 0.2713 & 0.0100 \\
$Z$\tablenotemark{g} & 0.0099  & 0.0020  & 0.00319 & 0.00064 & 0.01373 & 0.00270 & 0.0199 & 0.0040 & 0.0145 & 0.0022 \\
$O$\tablenotemark{g} & 0.00460 & 0.00069 & 0.00170 & 0.00026 & 0.00584 & 0.00093 & 0.0082 & 0.0013 & 0.0057 & 0.0008 \\
\enddata
\tablenotetext{a} {Gaseous content for the \ion{H}{2} regions, with the 
exception of O and C where the fractions of these elements 
trapped in dust grains have been  included, see text.}
\tablenotetext{b} {This paper, values for $t^2$ = 0.033.}
\tablenotetext{c} {\citet{duf82,pei00,rel02}, values for $t^2$ = 0.022.}
\tablenotetext{d} {\citet{pei93z,est98,est02}, values for $t^2$ = 0.024.}
\tablenotetext{e} {\citet{pei92,est99a,pei02a,est02}, values for $t^2$ =
0.037.}
\tablenotetext{f} {\citet{chr98,gre98,all01,all02,hol01}.}
\tablenotetext{g} {By mass.}
\end{deluxetable}

\clearpage

\begin{deluxetable}{l@{\hspace{24pt}}cc@{\hspace{12pt}}cc}
\tablewidth{0pt}
\tablecaption{Values of $\Delta Y/\Delta O$ and $\Delta Y/\Delta Z$
\label{tdy/dz}}
\tablehead{
\colhead{} &
\multicolumn{2}{c}{$\Delta Y$/$\Delta O$} & 
\multicolumn{2}{c}{$\Delta Y$/$\Delta Z$} \\
\cline{2-3} \cline{4-5}
\colhead{Object} &
\colhead{$t^2 = 0.00$\tablenotemark{a}} &
\colhead{$t^2 > 0.00$\tablenotemark{b}} &
\colhead{$t^2 = 0.00$\tablenotemark{a}} &
\colhead{$t^2 > 0.00$\tablenotemark{b}} }
\startdata
30 Dor -  $Y_p$(+Hc)                & $2.83 \pm 0.82$  &  $2.65 \pm 0.78$  &  $1.33 \pm 0.36$  &  $1.23 \pm 0.36$ \\
30 Dor -  NGC~346\tablenotemark{c}  & $7.27 \pm 2.10$  &  $3.48 \pm 1.01$  &  $3.05 \pm 0.90$  &  $1.51 \pm 0.46$ \\
M17\tablenotemark{d} -  $Y_p$(+Hc)  & $6.60 \pm 1.25$  &  $3.45 \pm 0.65$  &  $2.15 \pm 0.45$  &  $1.46 \pm 0.32$ \\
Sun - $Y_p$(+Hc)                    & $4.18 \pm 1.42$  &  $5.77 \pm 1.92$  &  $1.64 \pm 0.56$  &  $2.27 \pm 0.77$ \\
\enddata
\tablenotetext{a}{$Y_p$(+Hc,$t^2 = 0.00$) = 0.2475, \citet*{pei02z}.}
\tablenotetext{b}{$Y_p$(+Hc,$t^2 > 0.00$) = 0.2384, \citet{pei02z}.}
\tablenotetext{c}{\citet*{pei00}.}
\tablenotetext{d}{\citet{pei02a}.}
\end{deluxetable}

\clearpage

\begin{figure}
\begin{center}
\includegraphics[scale=0.77]{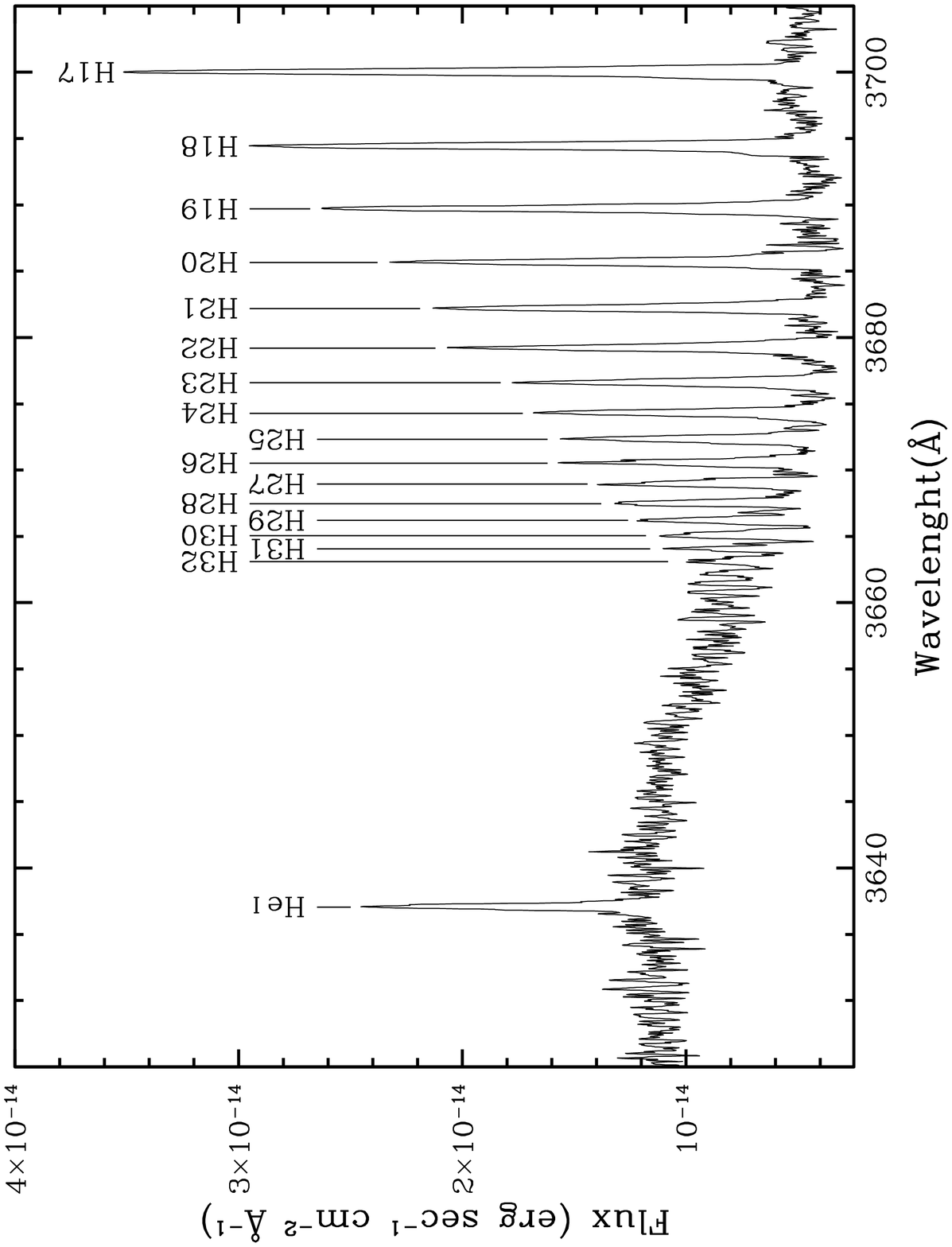}
\end{center}
\end{figure}

\figcaption[f1.eps]{
\label{fbalmer}
Section of the echelle spectrum at the Balmer limit region (observed fluxes).
}

\begin{figure}
\begin{center}
\includegraphics[scale=0.77]{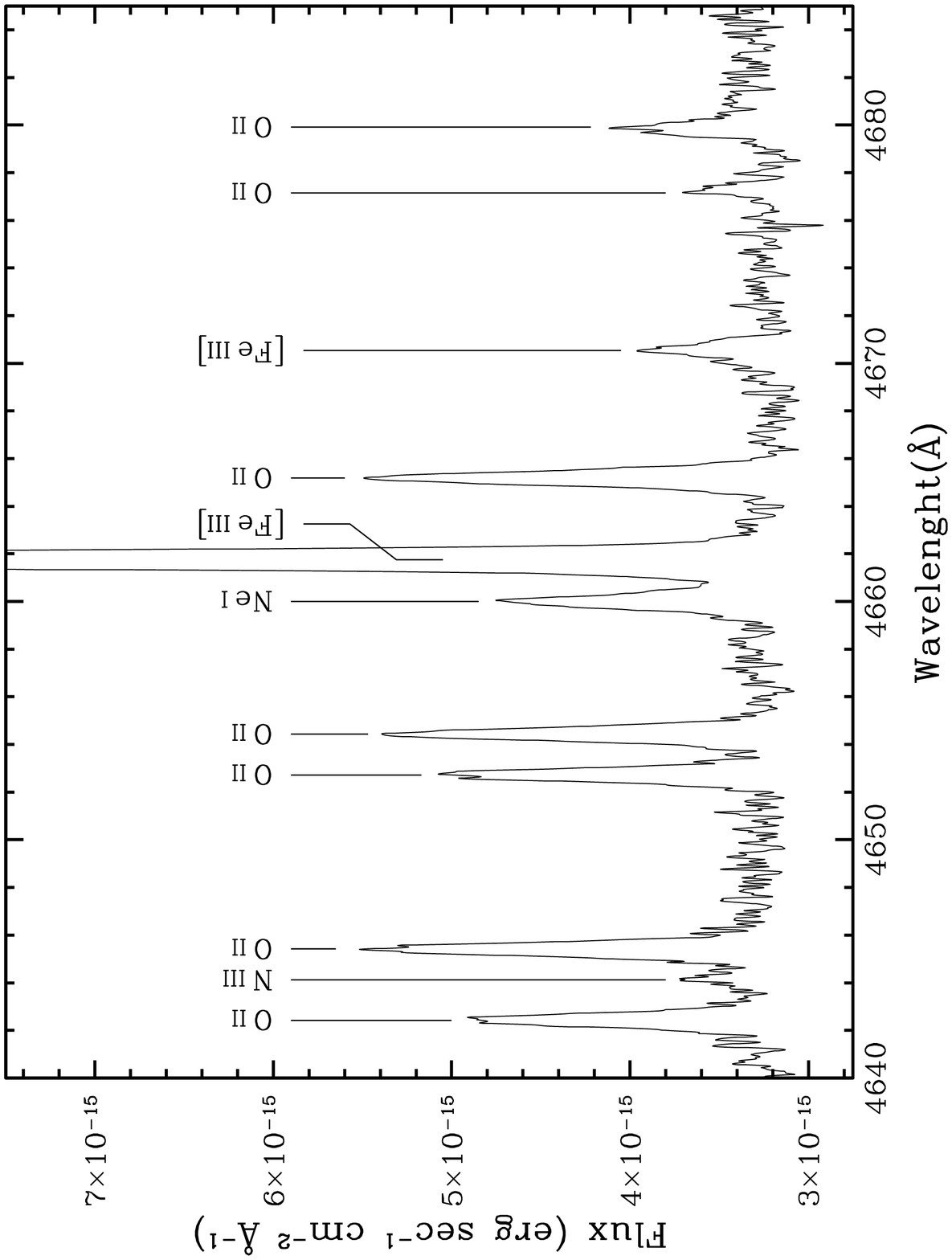}
\end{center}
\end{figure}

\figcaption[f2.eps]{
\label{foii}
Section of the echelle spectrum showing the individual emission 
lines of multiplet 1 of \ion{O}{2} (observed fluxes).
}

\end{document}